\documentclass{lmcs}
\pdfoutput=1
\usepackage[utf8]{inputenc}

\usepackage{lastpage}
\lmcsdoi{18}{3}{6}
\lmcsheading{}{\pageref{LastPage}}{}{}%
{Oct.~06,~2021}{Jul.~28,~2022}{}


\usepackage{amsmath,amssymb,amsfonts,amsthm}
\usepackage{graphicx}
\usepackage{algorithm}
\usepackage[noend]{algpseudocode}
\usepackage{multicol}
\usepackage{bm}
\usepackage{enumitem}
\usepackage{hyperref}

\newcommand{\N}{\mathbb N}
\newcommand{\R}{\mathbb R}
\newcommand{\Q}{\mathbb Q}
\newcommand{\eps}{\varepsilon}
\newcommand{\ff}{\mathcal F}
\newcommand{\fuse}{\sim}
\newcommand{\lsub}[1]{\mathrel{<_{#1}}}
\newcommand{\tok}[1]{\mathrel{\to_{#1}}}
\newcommand{\hmax}{h_{\mathrm{max}}}
\newcommand{\defisymbol}{\quad \bm{:=}\quad }
\newcommand{\TS}{\textsc{TameSucc}}
\newcommand{\TSs}{T}                        
\newcommand{\succe}{\textsc{Succ}}
\newcommand{\MT}{M}
\newcommand{\ISF}{\textsc{IsFusible}}
\newcommand{\ACAzero}{\mathrm{ACA}_0}

\DeclareMathOperator{\ord}{ord}
\DeclareMathOperator{\fus}{fus}

\newcommand{\exc}{\chi}

\begin{document}

\title{Fusible numbers and Peano Arithmetic}
\author[J. Erickson]{Jeff Erickson\lmcsorcid{0000-0002-5253-2282}}[a]
\author[G. Nivasch]{Gabriel Nivasch\lmcsorcid{0000-0001-5960-4253}}[b]
\author[J. Xu]{Junyan Xu\lmcsorcid{0000-0002-3789-2319}}[c]

\address{University of Illinois}
\email{jeffe@illinois.edu}

\address{Ariel University, Israel}
\email{gabrieln@ariel.ac.il}

\address{Cancer Data Science Laboratory, CCR, NCI, NIH}
\email{junyanxu.math@gmail.com}

\begin{abstract}
Inspired by a mathematical riddle involving fuses, we define the \emph{fusible numbers} as follows: $0$ is fusible, and whenever $x,y$ are fusible with $|y-x|<1$, the number $(x+y+1)/2$ is also fusible. We prove that the set of fusible numbers, ordered by the usual order on $\R$, is well-ordered, with order type $\eps_0$. Furthermore, we prove that the density of the fusible numbers along the real line grows at an incredibly fast rate: Letting $g(n)$ be the largest gap between consecutive fusible numbers in the interval $[n,\infty)$, we have $g(n)^{-1} \ge F_{\eps_0}(n-c)$ for some constant $c$, where $F_\alpha$ denotes the fast-growing hierarchy.

Finally, we derive some true statements that can be formulated but not proven in Peano Arithmetic, of a different flavor than previously known such statements: PA cannot prove the true statement ``For every natural number $n$ there exists a smallest fusible number larger than $n$.'' Also, consider the algorithm ``$M(x)$: if $x<0$ return $-x$, else return $M(x-M(x-1))/2$.'' Then $M$ terminates on real inputs, although PA cannot prove the statement ``$M$ terminates on all natural inputs.''
\end{abstract}

\maketitle

\section{Introduction}

A popular mathematical riddle goes as follows: We have available two fuses, each of which will burn for one hour when lit. How can we use the two fuses to measure 45 minutes? The fuses do not burn at a uniform rate, so we cannot predict how much of a fuse will burn after any period of time smaller than an hour.

The solution is to light both ends of one fuse simultaneously, and at the same time light one end of the second fuse. The first fuse will burn after 30 minutes. At that moment we light the second end of the second fuse. The second fuse will burn after an additional 15 minutes.

This riddle appears for example in Peter Winkler's puzzle collection \cite[p.~2]{winkler}, where he attributes it to Carl Morris. It was also aired by Ray Magliozzi in his \emph{Car Talk} radio show~\cite{cartalk}. Frank Morgan called it an ``old challenge'' in 1999~\cite{mathchat}. The puzzle (or variants using strings, shoelaces, or candles) even became a standard interview question at some job interviews.

Based on this puzzle, we define the set $\ff\subset\Q$ of \emph{fusible numbers} as the set of all times that can be similarly measured by using any number of fuses that burn for unit time. Formally, let us first define 
\begin{equation*}
x\fuse y=(x+y+1)/2\qquad\text{for $x,y\in \R$.}
\end{equation*}
If $|y-x|<1$,  then the number $x\fuse y$ (read ``$x$ fuse $y$'') represents the time at which a fuse burns, if its endpoints are lit at times $x$ and $y$, respectively. Then $\ff$ is defined recursively by $0\in\ff$, and $x\fuse y\in\ff$ whenever $x,y\in\ff$ with $|y-x|<1$.

Note that we do not allow lighting a fuse at only one endpoint. This entails no loss of generality, since the same effect can be obtained by lighting a fuse at both endpoints simultaneously, and then lighting a second fuse at both endpoints simultaneously. Hence, for example, $1\in\ff$ since $1 = (0\fuse 0)\fuse(0\fuse 0)$.

Every $z\in \ff$ is a nonnegative dyadic rational, meaning a nonnegative rational number whose denominator, when in lowest terms, is a power of $2$.

There might be several ways of obtaining a given fusible number $z\in\ff$. For example, $3/4\fuse 3/4=1/2\fuse 1=5/4$.

\begin{figure}
\centerline{\includegraphics{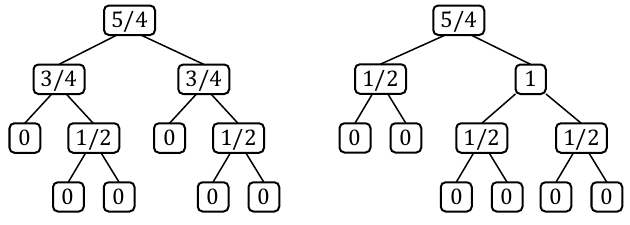}}
\caption{\label{fig_trees}Two different representations of the fusible number $5/4$.}
\end{figure}

We  distinguish between fusible numbers, which belong to $\Q$, and their \emph{representations}. A representation is an algebraic expression using ``$0$'' and ``$\fuse$" that gives rise to a fusible number. We can think of a representation as a full binary tree (i.e. a binary tree each of whose nodes has either zero or two children). Each leaf of the tree corresponds to a $0$, and each internal node corresponds to a $\fuse$ operation. Not every full binary tree is a valid representation, however, because of the $|y-x|<1$ requirement. Hence, $5/4\in\ff$ has the two representations
\begin{align*}
T_1 &=``(0\fuse (0\fuse 0))\fuse (0\fuse (0\fuse 0))",\\
T_2&=``(0\fuse 0)\fuse((0\fuse 0)\fuse(0\fuse 0))",
\end{align*}
whose corresponding trees are shown in Figure~\ref{fig_trees}. (When necessary, we distinguish representations from their corresponding fusible numbers by enclosing representations in quotation marks.)

We will usually represent fusible numbers with lowercase letters and representations with uppercase letters. The fusible number corresponding to a representation $T$ is denoted by $v(T)$.

The \emph{height} $h(T)$ of a representation $T$ is the height of the corresponding tree, i.e.~the length of the longest path from the root to a leaf. Hence, $h(``0")=0$ and $h(``0\fuse 0")=1$. Both representations $T_1$, $T_2$ of $5/4$ above have height $3$. The fusible number $11/8$ has a representation of height $3$ (as $11/8=3/4\fuse 1$), as well as one of height $4$ (as $11/8=7/8\fuse 7/8$ or $11/8=1/2\fuse 5/4$).

We have $v(T) = \sum_x 2^{-\mathrm{depth}(x)-1}$, where the sum ranges over all internal nodes $x$ of $T$.

\begin{figure}
\centerline{\includegraphics{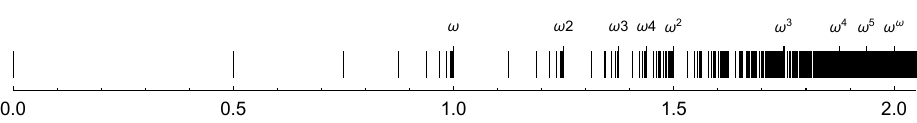}}
\caption{\label{fig_fusible}The fusible numbers and their corresponding ordinals.}
\end{figure}

There are infinitely many fusible numbers smaller than $1$. In fact, the fusible numbers smaller than $1$ are exactly those of the form $1-2^{-n}$ for\footnote{We denote $\N = \{0,1,2,3,\ldots\}$ and $\N^+=\{1,2,3,\ldots\}$.} $n\in\N$, given by $1-2^{-(n+1)} = 0\fuse(1-2^{-n})$. This sequence converges to $1$. The number $1$ itself is also fusible, given by $1=1/2\fuse 1/2$. Figure~\ref{fig_fusible} shows part of the set $\ff$.

\subsection{Our results}

As we show in this paper, the set $\ff$ of fusible numbers is related to ordinals, proof theory, computability theory, and fast-growing functions (in fact, functions growing much faster than Ackermann's function). In this paper we prove the following:

\begin{thm}\label{thm_eps0}
The set $\ff$ of fusible numbers, when ordered by the usual order ``$<$'' on $\R$, is well-ordered, with order type $\eps_0$.
\end{thm}

\begin{thm}\label{thm_fast}
The density of the fusible numbers along the real line grows very fast: Let $g(n)$ be the largest gap between consecutive fusible numbers in $\ff\cap [n,\infty)$. Then $g(n)^{-1} \ge F_{\eps_0}(n-7)$ for all $n\ge 8$, where $F_\alpha$ denotes the fast-growing hierarchy.
\end{thm}

\paragraph{Peano Arithmetic} After G\"odel published his incompleteness results, mathematicians became interested in finding examples of \emph{natural} true statements that cannot be proven in different formal theories. Peano Arithmetic is a nice example of a simple theory, since it allows making ``finitary'' statements (statements involving finite sets and structures built on the natural numbers) though not ``infinitary'' statements (statements involving infinite sets). Despite its limitations, a good portion of modern mathematics can be carried out in PA (see e.g.~Bovykin~\cite{bovykin}).

There are several known natural true statements that can be formulated in PA, but whose proofs require ``infinitary'' arguments and thus cannot be carried out in PA. The most famous examples concern Goodstein sequences (Kirby and Paris~\cite{KP}), hydras (Kirby and Paris~\cite{KP}), and a variant of Ramsey's theorem (Paris and Harrington~\cite{PH}). Further examples include another variant of Ramsey's theorem involving regressive colorings (Kanamori and McAloon~\cite{KMcA}); so-called worms (Beklemishev~\cite{beklemishev}, previously discovered by Hamano and Okada~\cite{hamano_okada}); a finite miniaturization of Kruskal's tree theorem (Friedman, mentioned in~\cite{simpson_trees}); a finite miniaturization of the Robertson--Seymour graph minor theorem (Friedman et al.~\cite{FRS}); and a finite miniaturization of a topological result of Galvin and Prikry (Friedman et al.~\cite{FMcAS}, also mentioned in~\cite{CS}). See the survey by Simpson~\cite{simpson_unprovable}.

There are many articles on the topic aimed at a wide readership. See for example Sladek~\cite{sladek} and Smory\'nski~\cite{smo1,smo2}.

As we show in this paper, fusible numbers are a source of a few other true statements that can be formulated but not proven in PA. Interestingly, these statements are of a different flavor than the previously known ones:

\begin{thm}\label{thm_PA}\ 
\begin{enumerate}
\item PA cannot prove the true statement ``For every $n\in\N$ there exists a smallest fusible number larger than $n$.''
\item PA cannot prove the true statement ``Every fusible number $x\in\ff$ has a maximum-height representation $\hmax(x)$.''
\item Consider the algorithm ``$\MT(x)$: if $x<0$ return $-x$, else return $\MT(x-\MT(x-1))/2$.'' Then $\MT$ terminates on real inputs, although PA cannot prove the statement ``$\MT$ terminates on all natural inputs.''
\end{enumerate}
\end{thm}

\paragraph{Organization of this paper} In Section~\ref{sec_basic} we derive some basic results about fusible numbers. Section~\ref{sec_ordinals} contains some necessary background on ordinals and fast-growing functions up to $\eps_0$. In Section~\ref{sec_tame} we prove the lower bound of Theorem~\ref{thm_eps0}, as well as Theorem~\ref{thm_fast}, by considering a subset of $\ff$ which we call \emph{tame} fusible numbers. In Section~\ref{sec_upper} we prove the upper bound of Theorem~\ref{thm_eps0}. In Section~\ref{sec_PA} we give some background on Peano Arithmetic and prove Theorem~\ref{thm_PA}. Section~\ref{sec_conj} presents a conjecture which, if true, would further clarify the structure of $\ff$. We conclude in Section~\ref{sec_discussion}.

Fusible numbers were first introduced by the first author at a talk in a satellite Gathering 4 Gardner event at UIUC at around 2010, in which he mentioned some of the above results. The slides of the talk are available at~\cite{jeffslides}. The third author later on released a manuscript~\cite{junyansurvey} with further results, as well as some corrections to the claims in the slides. The present paper builds upon these two unpublished works, also making some additional minor corrections.

\section{Basic results}\label{sec_basic}

In this section we derive some basic results about $\ff$. In particular, we give a simple proof that $\ff$ is well-ordered. As a corollary, we also derive an algorithm $\succe(x)$ for finding the smallest fusible number larger than $x$, as well as an algorithm $\ISF(x)$ for deciding whether $x\in\ff$ or not. The well-orderedness of $\ff$ implies that both algorithms terminate on all inputs.

\begin{obs}\label{obs_fuse}
Suppose $x\le y$ with $|y-x|<1$, and let $z=x\fuse y$. Then $x+1/2\le z<x+1$ and $y<z$.
\end{obs}

\begin{lem}[Window lemma]\label{lem_window}
Suppose $z\in\ff$, and let $0<m\le 1$. Then there exists a fusible number in the window $I=(z+1-2m,z+1-m]$.
\end{lem}

\begin{proof}
Let $z_n = z + 1 - 2^{-n}$ for $n\in\N$. Since $z_0 = z$ and $z_{n+1} = z\fuse z_n$, we have $z_n\in\ff$ for all $n$. One of these numbers belongs to $I$.
\end{proof}

\begin{lem}\label{lem_wo}
The set $\ff$, ordered by the usual order ``$<$" of real numbers, is well-ordered, meaning every nonempty subset of $\ff$ has a smallest element.
\end{lem}

\begin{proof}
Suppose for a contradiction that $\ff$ is not well-ordered, so there exist nonempty subsets $\mathcal G\subseteq\ff$ without a smallest element. Every such subset contains an infinite descending sequence $S = (z_1>z_2>z_3>\cdots)$.\footnote{To avoid relying on the axiom of choice, we can choose each $z_n$ canonically from $z_{n-1}$ in some way. For example, we could let $z_n$ be the element of $\mathcal G\cap (0,z_{n-1})$ with smallest denominator, and from among them, the one with the smallest numerator.} Each infinite descending sequence $S$ of $\ff$ converges to some limit $\lim S\in \R$. Let $\mathcal L$ be the set of all such limits. Then $\mathcal L$ must have a smallest element $\ell_0$, since the limit of an infinite descending sequence in $\mathcal L$ must also be in $\mathcal L$. So let $S_0=(z_1>z_2>z_3>\cdots)$ be an infinite descending sequence in $\ff$ converging to $\ell_0$.

Write $z_i = x_i\fuse y_i$ with $x_i\le y_i$. We can assume without loss of generality (by applying the infinite Ramsey theorem and taking a subsequence of $S_0$, if necessary) that the sequence $\{x_i\}$ is either ascending, descending, or constant. But by Observation~\ref{obs_fuse} and the minimality of $\ell_0$, the sequence $\{x_i\}$ cannot be descending. Hence, $\{x_i\}$ is either constant or ascending, and therefore $\{y_i\}$ is descending. Let $y^*$ be such that $x_1 \fuse y^*=\ell_0$. By Observation~\ref{obs_fuse} we have $y^*<\ell_0$, and by the continuity of ``$\fuse$" and the fact that $x_i\ge x_1$ for all $i$, we have $\lim y_i\le y^*$. This contradicts the minimality of $\ell_0$. 
\end{proof}

Given a well-ordered subset $\mathcal G\subset \R$ (with respect to the usual order ``$<$''), let $\ord(\mathcal G)$ be the ordinal corresponding to $\mathcal G$.\footnote{Note that $\mathcal G$ must be countable in order to be well-ordered, since after every element of $\mathcal G$ there must be an open interval disjoint from $\mathcal G$, which contains a rational number. Conversely, every countable ordinal $\alpha$ can be realized by some $\mathcal G\subset \R$.} Lemma~\ref{lem_wo} raises the problem of identifying the ordinal $\ord(\ff)$. As mentioned, we will prove that $\ord(\ff)=\eps_0$.

Given $r\in\R$, we will denote $\ord(\ff\cap [0,r))$ by $\ord(r)$. Conversely, given a countable ordinal $\alpha<\ord(\ff)$, let $\fus(\alpha)$ be the unique element $z\in\ff$ such that $\ord(z)=\alpha$. So for example, $\ord(0)=0$, $\ord(1/2)=1$, $\ord(3/4)=2$, $\ord(1)=\omega$, and $\ord(9/8)=\omega+1$. Conversely $\fus(\omega)=1$ and $\fus(\omega+1)=9/8$.

\begin{lem}\label{lem_closed}
$\ff$ is closed, meaning if $z_1<z_2<z_3<\cdots$ is a sequence in $\ff$ converging to $z\in\R$, then $z\in\ff$.
\end{lem}

\begin{proof}
Let $\mathcal L$ be the set of all ``missing limits'' in $\ff$, i.e. the set of all real numbers $x\notin\ff$ for which there exists an ascending sequence in $\ff$ converging to $x$. Assuming for a contradiction that $\mathcal L\neq\emptyset$, let $\ell=\inf\mathcal L$. We must have $\ell\in\mathcal L$, because otherwise we would obtain a contradiction to Lemma~\ref{lem_wo}.

Write $z_i = x_i\fuse y_i$. As before, we can assume without loss of generality that $\{x_i\}$ and $\{y_i\}$ are nondecreasing. Let $x = \lim x_i$ and $y=\lim y_i$. By the continuity of ``$\fuse$", we have $\ell=x\fuse y$. By Observation~\ref{obs_fuse} we have $\ell-1\le x\le \ell-1/2$. Hence, by the minimality of $\ell$, we must have $x\in\ff$. If $x>\ell-1$ then $y<\ell$, so by the minimality of $\ell$ we also have $y\in\ff$, implying that $\ell\in\ff$. And if $x=\ell-1$ (and so $y=\ell$), then $\ell=(x\fuse x)\fuse(x\fuse x)$.
\end{proof}

Hence, $\ff$ contains two types of nonzero fusible numbers: \emph{successors} and \emph{limits}. They correspond exactly to the successor and limit nonzero ordinals below $\ord(\ff)$.

Given a dyadic fraction $r= p/q$ with $p$ odd and $q=2^n$, we denote $e(r)=n$ and call it the \emph{exponent} of $r$.

\begin{obs}\label{obs_dyadic}
Let $r_1=(p-1)/q$ and $r_2=p/q$ be two dyadic fractions (not necessarily in lowest terms) with $q=2^n$. Then every dyadic fraction $r_1<r<r_2$ satisfies $e(r)>n$.
\end{obs}

\begin{obs}\label{obs_de}
Let $x,y,z$ be dyadic fractions with $z=x\fuse y$, and suppose that $1\le e(x)\le e(y)$. If $e(x)<e(y)$ then $e(z) = e(y)+1$. And if $e(x) = e(y)$ then $e(z)\le e(x)$.
\end{obs}

\begin{cor}\label{cor_de}
For every representation $T$ of a fusible number $z=v(T)$, we have $h(T) \ge e(z)$.
\end{cor}

\begin{lem}\label{lem_fwd}
Let $T$ be a representation of $z=v(T)$ with height $h(T)=n$. Then $z'=z+2^{-(n+1)}\in\ff$, and it has a representation of height $n+1$.
\end{lem}

\begin{proof}
By induction on $n$. If $n=0$ then $T=``0"$ and $z=0$, and then $z'=1/2=0\fuse 0$. Now suppose the claim is true for height $n-1$, and let $T=``T_1\fuse T_2"$ be a representation of height $n$. Then $h(T_1), h(T_2)\le n-1$, with equality in at least one case. Say $h(T_1)=n-1$. Let $x=v(T_1)$ and $y=v(T_2)$. We have $|y-x|<1$ and $e(x),e(y)\le n-1$, so actually $|y-x|\le 1-2^{-(n-1)}$. By the induction hypothesis, there exists a representation $T'$ of $x'=x+2^{-n}$ with $h(T')=n$. Furthermore, we still have $|y-x'|<1$, so $``T'\fuse T_2"$ is valid, and it represents $z'=z+2^{-(n+1)}$.
\end{proof}

\begin{lem}\label{lem_back}
Let $T$ be a representation of $z=v(T)$ with height $h(T)=n>0$. Then $z'=z-2^{-n}\in\ff$, and it has a representation of height at least $n-1$, as well as a representation of height at most $n$.
\end{lem}

\begin{proof}
Again by induction on $n$. If $n=1$ then $T=``0\fuse 0"$ and $z=1/2$, and then $z'=0$. Now suppose the claim is true for height $n-1$, and let $T=``T_1\fuse T_2"$ be a representation of height $n$. Then $h(T_1), h(T_2)\le n-1$, with equality in at least one case. Say $h(T_1)=n-1$. Let $x=v(T_1)$ and $y=v(T_2)$. We have $|y-x|<1$ and $e(x),e(y)\le n-1$, so actually $|y-x|\le 1-2^{-(n-1)}$. By the induction hypothesis, $x'=x-2^{-(n-1)}$ has a representation $T'$ with $h(T')\ge n-2$, as well as a representation $T''$ with $h(T'')\le n-1$. If we still have $|y-x'|<1$ then $``T'\fuse T_2"$ and $``T''\fuse T_2"$ are valid and we are done. Otherwise, we have $y-x'=1$, and so $z'=z-2^{-n}=y=x'+1$, so $T_2$ and $``(T'\fuse T')\fuse (T'\fuse T')"$ are two representations of $z'$. The first one has height at most $n-1$, and the second one has height at least $n$.
\end{proof}

Lemma~\ref{lem_fwd} implies that every fusible number has a maximum-height representation, since otherwise we would obtain an infinite decreasing sequence in $\ff$. Given $z\in\ff$, denote by $\hmax(z)$ the maximum of $h(T)$ over all representations $T$ with $v(T)=z$.

\begin{lem}\label{lem_recognize_limits}
Let $z\in\ff$ be a fusible number satisfying $\hmax(z)>e(z)$. Then $z$ is a limit.
\end{lem}

\begin{proof}
Assuming for a contradiction that $z$ is a successor, let $z'\in\ff$ be the predecessor of $z$. Let $n=\hmax(z)$. By Lemma~\ref{lem_back} we have $z''=z-2^{-n}\in \ff$. Hence, $z''\le z'<z$. Since $e(z'')=n$, by Observation~\ref{obs_dyadic} we have $e(z')\ge n$ as well. Hence, by Lemma~\ref{lem_fwd} we have $z'''=z'+2^{-(n+1)}\in\ff$, and by Observation~\ref{obs_dyadic} we still have $z'''<z$. This contradicts the fact that $z'$ is the predecessor of $z$.
\end{proof}

By the continuity of ``$\fuse$'', if $z=x\fuse y$ where $x,y,z\in\ff$ and one of $x$, $y$ is a limit, then $z$ is also a limit. The converse is also true:

\begin{lem}\label{lem_lim_from_lim}
Let $z>1$ be a limit fusible number. Then $z$ can be written as $z=x\fuse y$ where $x,y\in\ff$ and at least one of $x$, $y$ is a limit.\footnote{Lemmas~\ref{lem_lim_from_lim}, \ref{lem_get_conv}, \ref{lem_recognize_limits2} and Corollary~\ref{cor_recognize_lim_together} are not used in the rest of this paper.}
\end{lem}

\begin{proof}
Let $\{z_i\}$ be an increasing sequence converging to $z$. Let $z_i=x_i\fuse y_i$, and assume without loss of generality that each of $\{x_i\}$, $\{y_i\}$ is either constant or increasing. At least one of these sequences must be increasing. Let $x=\lim x_i$ and $y=\lim y_i$. We have $x,y\in\ff$ by Lemma~\ref{lem_closed}. If $x,y<z$, then $z=x\fuse y$, and we are done. Otherwise, suppose for example that $x=z-1$ and $y=z$. Then, $z=(x\fuse x)\fuse(x\fuse x)$. In this case, $x\fuse x$ is a limit by Observation~\ref{obs_de} and Lemma~\ref{lem_recognize_limits}, and we are done again.
\end{proof}

Since $\ff$ is well-ordered, every fusible number has a successor. Denote the successor of $z\in\ff$ by $s(z)$.

\begin{lem}\label{lem_hmax_succ}
Let $z\in\ff$ be a fusible number with $\hmax(z)=n$. Then its successor is $s(z)=z+2^{-(n+1)}$, and its maximum height is $\hmax(s(z))=n+1$.
\end{lem}

\begin{proof}
We already know by Lemma~\ref{lem_fwd} that $z'=z+2^{-(n+1)}\in\ff$. By Corollary~\ref{cor_de} we have $e(z) \le n$, so $e(z') = n+1$. Suppose for a contradiction that there exist fusible numbers $z''\in(z,z')$. Let $m$ be the smallest height of a representation of such a fusible number. Choose $z''\in\ff\cap (z,z')$ minimally from among those having a representation $T$ of height $m$. By Corollary~\ref{cor_de} and Observation~\ref{obs_dyadic} we have $m\ge e(z'') \ge n+2$. Hence, by Lemma~\ref{lem_back} and Observation~\ref{obs_dyadic}, there exists $z'''\in\ff\cap [z,z'')$ with a representation of height at least $n+1$ and a representation of height at most $m$. If $z'''>z$ this contradicts the minimality of $m$ or $z''$, whereas if $z'''=z$ this contradicts the assumption $\hmax(z)=n$. Hence, we must have $\ff\cap(z,z') = \emptyset$, so $s(z)=z'$. If $\hmax(z')$ were larger than $n+1$, then again applying Lemma~\ref{lem_back} on $z'$ would lead to a contradiction.
\end{proof}

\begin{cor}\label{cor_next_omega}
Let $z\in\ff$ have maximum height $\hmax(z)=n$. Then the next $\omega$ fusible numbers following $z$ are $z+2^{-n} - 2^{-(n+m)}$ for $m=1,2,3,\ldots$. 
\end{cor}

\begin{lem}\label{lem_pred}
Let $z=p/q\in\ff$ be a successor fusible number, with $p$ odd. Then its predecessor is $z'=(p-1)/q$.
\end{lem}

\begin{proof}
By Lemma~\ref{lem_recognize_limits} we have $\hmax(z)=e(z)=n$ where $q=2^n$. Hence, by Lemma~\ref{lem_back} we know that $z'\in\ff$. Let $z''$ be the predecessor of $z$, and suppose for a contradiction that $z'<z''<z$. Then Observation~\ref{obs_dyadic} together with Lemma~\ref{lem_hmax_succ} yields a contradiction.
\end{proof}

\begin{lem}\label{lem_get_conv}
Let $z\in\ff$ be a limit, and let $m=e(z)$. Then $z_n=z-2^{-(m+n)}\in\ff$ for all $n\in\N$.
\end{lem}

\begin{proof}
Given $n$, suppose for a contradiction that $z_n\notin\ff$. Since $\ff$ is closed, there exists a maximum element $z'\in\ff\cap[0,z_n)$. Let $z''=s(z')$. Since $z$ is a limit, we must have $z_n<z''<z$. By Observation~\ref{obs_dyadic} on $z_n$ and $z$, we have $e(z'')>e(z_n)$. But $z''$ is a successor ordinal, so by Lemma~\ref{lem_pred} and Observation~\ref{obs_dyadic} on $z'$ and $z''$, we have $e(z_n)> e(z'')$. Contradiction.
\end{proof}

Finally, we prove the converse of Lemma~\ref{lem_recognize_limits}.

\begin{lem}\label{lem_recognize_limits2}
Suppose $z\in\ff$ is a limit. Then $\hmax(z) > e(z)$.
\end{lem}

\begin{proof}
Let $n=e(z)$. In light of Corollary~\ref{cor_de} and Lemma~\ref{lem_hmax_succ}, it is enough to prove the existence of a fusible number $z'\in (z,z+2^{-(n+1)})$. If $z=1$ then we can take $z'=9/8\in\ff$ and we are done. Otherwise, by Lemma~\ref{lem_lim_from_lim}, we have $z=x\fuse y$ where $y$ is a limit. Denote $n_1=e(x)$, $n_2=e(y)$.

Suppose first that $n_1\le n_2$. Then $n\le n_2+1$. Since $|y-x|<1$, we actually have $|y-x|\le 1-2^{-n_2}$. By structural induction, we can assume that there exists a fusible number $y'\in (y, y+2^{-(n_2+1)})$. Hence, $z' = x\fuse y'$ is valid, and it is smaller than $z+2^{-(n_2 + 2)}\le z+2^{-(n+1)}$, as desired.

Otherwise, we have $n_1>n_2$ and $x$ is a successor fusible number. In this case we have $n = n_1+1$, and $|y-x|\le 1-2^{-n_1}$. By Lemma~\ref{lem_recognize_limits} we have $\hmax(x) = n_1$, so By Lemma~\ref{lem_fwd} we have $x'=x + 2^{-(n_1+1)}\in\ff$. Furthermore, by Lemma~\ref{lem_get_conv} we have $y'=y - 2^{-(n_1+2)}\in\ff$. Hence, $z'=x'\fuse y'$ is valid, and it equals $z+ 2^{-(n_1+3)} = z + 2^{-(n+2)}$.
\end{proof}

\begin{cor}\label{cor_recognize_lim_together}
Let $z\in\ff$. Then $z$ is a limit if and only if $\hmax(z)>e(z)$.
\end{cor}

\subsection{Algorithms}

\begin{table}
\begin{multicols}{2}
\begin{algorithm}[H]
\caption{Finds $\min(\ff\cap(r,\infty))$}
\begin{algorithmic}[5]
\Procedure{Succ}{$r$}
\If{$r<0$}
\State\Return{$0$}
\EndIf
\State $min\gets\infty$
\State $y\gets r$
\Repeat
\State $x\gets$\Call{Succ}{$2r-y-1$}
\State $y\gets$\Call{Succ}{$2r-x-1$}
\If{$x\fuse y<min$}
\State $min\gets x\fuse y$
\EndIf
\State $y\gets y - 1/$\Call{Denominator}{$y$}
\Until{$y\le r-1/2$}
\State\Return{$min$}
\EndProcedure
\end{algorithmic}
\end{algorithm}

\begin{algorithm}[H]
\caption{Decides whether $r\in\ff$}
\begin{algorithmic}[5]
\Procedure{IsFusible}{$r$}
\If{$r$ is not a dyadic rational}
\State\Return{false}
\EndIf
\State\Return{$r=$\Call{WeakSucc}{$r$}}
\EndProcedure
\end{algorithmic}
\end{algorithm}

\begin{algorithm}[H]
\caption{Finds $\min(\ff\cap[r,\infty))$}
\begin{algorithmic}[5]
\Procedure{WeakSucc}{$r$}
\If{$r\le 0$}
\State\Return{$0$}
\EndIf
\State $min\gets\infty$
\State $y\gets r$
\Repeat
\State $x\gets$\Call{WeakSucc}{$2r-y-1$}
\If{$x=2r-y-1$}
\State \Return{$r$}
\EndIf
\State $y\gets$\Call{WeakSucc}{$2r-x-1$}
\If{$y=2r-x-1$}
\State \Return{$r$}
\EndIf
\If{$x\fuse y<min$}
\State $min\gets x\fuse y$
\EndIf
\State $y\gets y - 1/$\Call{Denominator}{$y$}
\Until{$y<r-1/2$}
\State\Return{$min$}
\EndProcedure
\end{algorithmic}
\end{algorithm}
\end{multicols}
\caption{\label{table_algs}Some algorithms for fusible numbers.}
\end{table}

Suppose we are given $z\in\ff$ and we want to find $s(z)$, or we want to decide whether a given number $z\in\Q$ belongs to $\ff$. It is not clear a priori that these problems are computable, since there are usually infinitely many fusible numbers smaller than $z$, all of which might potentially be useful in building $z$ or fusible numbers larger than~$z$.

Table~\ref{table_algs} shows a recursive algorithm $\succe(r)$ for the slightly more general problem of, given $r\in\Q$, finding the smallest fusible number $z$ that is strictly larger than $r$. It also shows a decision algorithm $\ISF$, which relies on $\textsc{WeakSucc}$, a modification of $\succe$.

\begin{lem}
Algorithms $\succe$, $\textsc{WeakSucc}$, and $\ISF$ terminate on all inputs and return the correct output.
\end{lem}

\begin{proof}
Consider algorithm $\succe$. Given $r\in\Q$, let $z=s(r)$ be the smallest fusible number larger than $r$. If $r<1/2$ the claim is easy to verify, so assume $r\ge 1/2$. We must have $z=x\fuse y$ for some $x,y\in\ff$ satisfying $r-1<x\le y\le r$ and $y>r-1/2$. We will show that the algorithm tries all the relevant candidates for $x,y$ by increasing value of $x$ and decreasing value of $y$, and that there is a finite number of such candidates.

The following invariant is true at the beginning of the loop (line $6$): The variable $y$ holds either $r$ or a fusible number smaller than $r$, and if $y<r$ then the variable $\mathit{min}$ holds the smallest fusible number $z>r$ of the form $z=x'\fuse y'$ with $y<y'\le r$. Then, line $7$ increases $x$ to the next possible candidate, and line $8$ sets $y$ to the fusible number that minimizes $z'=x\fuse y$ subject to $z'>r$.

Line $8$ never sets $y$ larger than its previous value $y_\mathrm{old}$: We have $2r-x-1 < 2r - (2r-y_\mathrm{old} - 1)-1 = y_\mathrm{old}$, so if $y_\mathrm{old}<r$ then $y_\mathrm{old}$ is fusible, and hence (assuming that the recursive call in line $8$ returns the correct value) line 8 sets $y$ to $s(2r-x-1) \le y_\mathrm{old}$. Further, as we prove below, the recursive call on line $8$ never returns a value larger than $r$.

In addition, by Lemma~\ref{lem_pred}, line $11$ replaces the successor fusible number $y$ by its predecessor, so the algorithm does not miss any candidate. 

We still have to prove that the recursive calls on lines $7$, $8$ return the correct values $s(2r-y-1)$, $s(2r-x-1)$. This will follow by ordinal induction once we show that these values are smaller than $z$. For this, first note that at the beginning of the loop we always have $y>r-1/2$, which implies $2r-y-1<r-1/2$. Since every interval of length $1/2$ contains a fusible number, we have $s(2r-y-1)<r$. Hence, by ordinal induction line $7$ indeed sets $x$ to $s(2r-y-1)$.

Next, recall that at the beginning of the loop we always have $y\le r$. Hence, after line $7$ we have $x> 2r-y-1\ge r-1$. Denote $m=x+1-r$, so $0<m<1$. By the window lemma (Lemma~\ref{lem_window}) there exists a fusible number in the interval $(x+1-2m,x+1-m]$. But $x+1-2m = 2r-x-1$ and $x+1-m=r$. Hence, by ordinal induction line $8$ correctly sets $y$ to $s(2r-x-1)$, which is at most $r$.

Finally, the loop must end after a finite number of iterations, or else we would have an infinite descending sequence of fusible numbers.

The proof for the other two algorithms is similar. Lines $2$, $3$ in $\ISF$ are only for efficiency purposes.
\end{proof}

In algorithm $\succe$, the first candidate $x\fuse y$ is not always the best one. Consider for example $r = 33/16\in\ff$. Its successor is $r+1/4096 = 8449/4096=19/16\fuse 3969/2048$, but the first candidate $\succe$ tries is $r+1/2048 = 9/8\fuse 2049/1024$.

\paragraph{Running time} We will prove below that, already for $r\in\N$, algorithm $\succe(r)$ has an incredibly large running time in the size of the input. On the other hand, we have not been able to rule out the possibility that algorithms $\textsc{WeakSucc}$ and $\ISF$ run in polynomial time in the size of the input. These two latter algorithms terminate very quickly when $r\in\N$.

\section{Background on ordinals}\label{sec_ordinals}

The ordinal $\eps_0$ is the limit of $\omega,\omega^\omega,\omega^{\omega^\omega}, \ldots$. Each ordinal $\alpha<\eps_0$ can be uniquely written in \emph{Cantor Normal Form} (CNF) as $\alpha = \omega^{\alpha_1} + \cdots + \omega^{\alpha_k}$, for some $k\ge 0$ and some ordinals $\alpha_1, \ldots, \alpha_k$ satisfying $\alpha>\alpha_1\ge\alpha_2\ge\cdots\ge\alpha_k$. Alternatively, $\alpha$ can be uniquely written in \emph{weighted CNF} as $\alpha = \omega^{\alpha_1}n_1 + \cdots + \omega^{\alpha_k}n_k$, for some $k\ge 0$, some ordinals $\alpha>\alpha_1>\cdots>\alpha_k$, and some positive integers $n_1, \ldots, n_k$.

Given a limit ordinal $\alpha<\eps_0$, the \emph{canonical sequence} $[\alpha]_n$ for $n\ge 1$ is a sequence that converges to $\alpha$. It is defined as follows: For limit $\beta<\omega^{\alpha+1}$ we let $[\omega^\alpha +\beta]_n = \omega^\alpha+[\beta]_n$. We let $[\omega^{\alpha+1}]_n = \omega^\alpha n$. And for limit $\alpha$ we let $[\omega^\alpha]_n = \omega^{[\alpha]_n}$. For example, $[\omega^{\omega+2}7]_5 = \omega^{\omega+2}6 + \omega^{\omega+1}5$.

Let us define for successor ordinals $[\alpha+1]_n = \alpha$, independently of $n$. A \emph{canonical descent} is a sequence $\alpha_1 > \alpha_2 > \cdots > \alpha_k$ in which $\alpha_{i+1} = [\alpha_i]_{n_i}$ for each $i$, for some parameters $n_i$. We call the parameters $n_i$ the \emph{descent parameters}.

For each pair of ordinals $\beta<\alpha<\eps_0$ there exists a canonical descent starting at $\alpha$ and ending at $\beta$. If there exists such a descent in which all the descent parameters are bounded by $k$, then we write $\alpha\tok{k}\beta$. Given $\alpha,\beta$, the shortest canonical descent from $\alpha$ to $\beta$ can be found greedily, by choosing at each step the smallest descent parameter that gives an ordinal at least $\beta$. This is implied by the following lemma:

\begin{lem}[Bachmann property~\cite{bachmann,rose}]\label{lem_nodetour}
Let $\alpha,\beta<\eps_0$ be limit ordinals satisfying $[\alpha]_n<\beta<\alpha$. Then $[\beta]_1\ge [\alpha]_n$ and $[\beta]_2>[\alpha]_n+1$.
\end{lem}

\begin{proof}
Without loss of generality we can assume $\alpha$ consists of a single term. Then we use induction on the height of $\alpha$.
\end{proof}

Hence, if $[\alpha]_n<\beta<\alpha$ (in particular, if $\beta=[\alpha]_m$ for some $m>n$), then every canonical descent from $\beta$ to $0$ must contain $[\alpha]_n$. Furthermore, if all the descent parameters are at least $2$, then the canonical descent must contain $[\alpha]_n+1$ as well.

We will need the following lemma:

\begin{lem}[Hiccup lemma]\label{lem_hiccup}
Let $\alpha,\beta<\eps_0$ be limit ordinals satisfying $\beta<\omega^\alpha$, and let $k\ge 2$. If $\beta = \omega^{[\alpha]_n+1}$ then we have $\bigl[\omega^{[\alpha]_n}+\beta\bigr]_k = \omega^{[\alpha]_n}+[\beta]_{k-1}$.\footnote{Note that if $\beta\ge \omega^{[\alpha]_n+1}$ then $\omega^{[\alpha]_n}+\beta=\beta$.} For every other $\beta$ we have $\bigl[\omega^{[\alpha]_n}+\beta\bigr]_k = \omega^{[\alpha]_n}+[\beta]_k$.
\end{lem}

\begin{proof}
If $\beta<\omega^{[\alpha]_n+1}$ then the expression ``$\omega^{[\alpha]_n}+\beta$'' is in CNF, so the claim follows by the definition of the canonical sequence. If $\beta = \omega^{[\alpha]_n+1}$ the claim is immediate. Hence, suppose $\beta>\omega^{[\alpha]_n+1}$, so $\omega^{[\alpha]_n}+\beta=\beta$. If $\beta$ consists of at least two terms then the first term is unchanged in $[\beta]_k$, and the claim follows. Finally, if $\beta = \omega^\gamma$ for some $[\alpha]_n<\gamma<\alpha$ then we consider separately the cases where $\gamma$ is a successor or a limit ordinal. We make use of the Bachmann property (Lemma~\ref{lem_nodetour}) in the latter case, using the fact that $k\ge 2$.
\end{proof}

Given $n\ge 0$, denote $\tau_n = \omega^{\omega^{\cdots^\omega}}$ with height $n$. In other words, $\tau_0 = 1$, and $\tau_{n+1} = \omega^{\tau_n}$ for $n\ge 0$.

\paragraph{Left-subtraction of ordinals} Given $\alpha\ge \beta$, define $-\beta+\alpha$ as the unique ordinal $\gamma$ such that $\beta+\gamma=\alpha$. In other words, $-\beta+\alpha = \ord(\alpha\setminus\beta)$ under von Neumann's definition of ordinals. For example, $-(\omega^\omega+\omega^3\cdot 2) + (\omega^\omega+\omega^4) = \omega^4$, and $-5+\omega=\omega$. In general, for every $\alpha<\omega^\beta$ we have $-\alpha+\omega^\beta=\omega^\beta$. Left-subtraction satisfies $-(\beta+\gamma)+\alpha = -\gamma+(-\beta+\alpha)$; and $-\beta+(\gamma+\alpha) = (-\beta+\gamma)+\alpha$ whenever $\gamma\ge \beta$.

\subsection{Natural sum and product}

Let $\alpha = \omega^{\alpha_1}+\cdots+\omega^{\alpha_n}$ and $\beta=\omega^{\beta_1} + \cdots+\omega^{\beta_m}$ be two ordinals in Cantor normal form, with $m,n\ge 0$ and $\alpha_1\ge \cdots\ge \alpha_n$ and $\beta_1\ge\cdots\ge \beta_m$. Then their \emph{natural sum} is given by
\begin{equation*}
\alpha\oplus\beta = \omega^{\gamma_1} + \cdots + \omega^{\gamma_{n+m}},
\end{equation*}
where $\gamma_1, \ldots, \gamma_{n+m}$ are $\alpha_1,\ldots,\alpha_n,\beta_1,\ldots,\beta_m$ sorted in nonincreasing order. For example, if $\alpha = \omega^3+\omega$ and $\beta=\omega^4+\omega+1$, then $\alpha\oplus\beta = \omega^4+\omega^3+\omega 2+1$ (whereas $\alpha+\beta=\omega^4+\omega+1$ and $\beta+\alpha = \omega^4+\omega^3+\omega$). The natural sum is commutative and associative.

Let $A$ and $B$ be two disjoint well-ordered sets, with respective well-orders $\lsub{A}$ and $\lsub{B}$. Define the partial order $\prec$ on $C=A\cup B$ by $x\prec y$ whenever $x,y\in A$ with $x\lsub{A} y$, or $x,y\in B$ with $x\lsub{B} y$. Let $\alpha$ and $\beta$ be the order types of $A$ and $B$, respectively. There might be several ways of \emph{linearizing} the partial order of $C$, i.e.~extending it into a total order. The largest possible order type of such a linearization equals $\alpha\oplus\beta$ (de Jongh and Parikh~\cite{dJP}).

Let $\alpha = \omega^{\alpha_1}+\cdots+\omega^{\alpha_n}$ and $\beta=\omega^{\beta_1} + \cdots+\omega^{\beta_m}$ as before. Then their \emph{natural product} is given by
\begin{equation*}
\alpha\otimes\beta = \bigoplus_{i,j} \omega^{\alpha_i\oplus\beta_j}.
\end{equation*}
The natural product is commutative and associative, and it distributes over the natural sum. If $\alpha>\beta$ then $\gamma\oplus \alpha > \gamma\oplus \beta$, and if $\gamma>0$ then also $\gamma\otimes\alpha > \gamma\otimes\beta$.

Let $A$ and $B$ be well-ordered sets, with respective well-orders $\lsub{A}$ and $\lsub{B}$. Define the partial order $\prec$ on $C=A\times B$ by $(a_1,b_1)\prec (a_2,b_2)$ whenever both $a_1\lsub{A} a_2$ and $b_1\lsub{B} b_2$. Then the largest possible order type of a linearization of $C$ equals $\alpha\otimes\beta$~\cite{dJP}.

\subsection{The fast-growing hierarchy}

Here we largely follow Buchholz and Wainer~\cite{BW}. The \emph{fast-growing hierarchy} is a family of functions $F_\alpha:\N^+\to\N^+$ indexed by ordinals $\alpha$, where $\alpha$ ranges up to some large countable ordinal. It has the property that for $\beta>\alpha$, the function $F_\beta(n)$ grows much faster than $F_\alpha(n)$. In fact, for every fixed $k$, the function $F_\beta(n)$ eventually overtakes $F_\alpha^{(k)}(n)$, where $f^{(k)} = f\circ f\circ\cdots\circ f$ denotes the $k$-fold composition of the function $f$.

The hierarchy up to $\eps_0$ is known as the \emph{Wainer hierarchy}. It is defined as follows:
\begin{align*}
F_0(n) &= n + 1,\\
F_{\alpha+1}(n) &= F_\alpha^{(n)}(n),\\
F_\alpha(n) &= F_{[\alpha]_n}(n),\qquad\text{for limit $\alpha<\eps_0$.}
\end{align*}
The well-known \emph{Ackermann function} $A(n)$ corresponds to $F_\omega$ in this hierarchy, in the sense that $F_\omega(n-c)\le A(n) \le F_\omega(n+c)$ for some constant $c$.

The fast-growing hierarchy is related to the \emph{Hardy hierarchy}, which is defined as follows:
\begin{align*}
H_0(n) &= n,\\
H_{\alpha+1}(n) &= H_\alpha(n+1),\\
H_\alpha(n) &= H_{[\alpha]_n}(n),\qquad\text{for limit $\alpha<\eps_0$.}
\end{align*}

\begin{lem}\label{lem_HF}
For every $\alpha<\eps_0$ and every $n$ we have $F_\alpha(n) = H_{\omega^\alpha}(n)$.
\end{lem}

\begin{proof}
It follows by ordinal induction on $\beta$ that for every $\alpha = \omega^{\alpha_1} + \cdots + \omega^{\alpha_k}$ in CNF and every $\beta<\omega^{\alpha_k+1}$, we have $H_{\alpha+\beta}(n) = H_\alpha(H_\beta(n))$. Then the claim follows by ordinal induction on $\alpha$.
\end{proof}

\begin{lem}\label{lem_H_basic}
Let $\alpha<\eps_0$. Then:
\begin{itemize}
\item $H_\alpha$ is strictly increasing.
\item Suppose $\alpha\tok{k}\beta$. Then $H_\alpha(n)> H_\beta(n)$ for every $n> k$.
\item The same properties hold for the functions $F_\alpha$.
\end{itemize}
\end{lem}

\begin{proof}
The first two claims are proven simultaneously by ordinal induction on $\alpha$. Then the third claim follows by Lemma~\ref{lem_HF}.
\end{proof}

Finally, we top the hierarchy by defining $F_{\eps_0}(n) = F_{\tau_n}(n)$. Hence, $F_{\eps_0}(n)$ eventually overtakes $F_\alpha(n)$ for every $\alpha<\eps_0$.

\begin{lem}\label{lem_columnize_eps0}
Let $z(n) = H_{\tau_n}(2)$. Then $z(n)>F_{\eps_0}(n-2)$ for every $n\ge 3$.
\end{lem}

\begin{proof}
We first observe the following: Suppose $\alpha\tok{2}\beta$, and let $r$ be the number of successor ordinals in the canonical descent from $\alpha$ to $\beta$ always using descent parameter $2$. Then $H_{\alpha}(n) \ge H_\beta(n+r)$ for every $n\ge 2$. Indeed, this follows by repeated application of the definition of $H$ and Lemma~\ref{lem_H_basic}.

Now consider the particular case $\alpha = \tau_{n+1}$, $\beta=\tau_n$. In this case, the canonical descent from $\alpha$ to $\beta$ contains at least $n$ successor ordinals: Consider the ordinals $\gamma_i = \delta_{i,i-1}$ for $1\le i\le n$, where $\delta_{i,0}=  \tau_{n-i+1}+1$ and $\delta_{i,j+1} = \omega^{\delta_{i,j}}$. (For example, for $n=3$ we have $\gamma_1 = \omega^{\omega^\omega}+1$, $\gamma_2 = \omega^{\omega^\omega+1}$, $\gamma_3 = \omega^{\omega^{\omega+1}}$.) Then the descent from $\alpha$ to $\beta$ contains $\gamma_{n}+1$ for each $i=n,\ldots,2$, as well as $\gamma_1$.

Hence, $z(n) = H_{\tau_n}(2) \ge H_{\tau_{n-1}}(n) = F_{\tau_{n-2}}(n) > F_{\tau_{n-2}}(n-2) = F_{\eps_0}(n-2)$.
\end{proof}

Hence, even though for fixed $\alpha<\eps_0$, the function $H_\alpha$ grows slower than $F_\alpha$, once we ``columnize'' by taking $\alpha = \tau_n$, the rates of growth of both functions match up.

\section{Lower bounds: Tame fusible numbers}\label{sec_tame}

In this section we prove the lower bound $\ord(\ff)\ge \eps_0$ of Theorem~\ref{thm_eps0}, as well as Theorem~\ref{thm_fast}. For this we consider a subset $\ff'\subset\ff$ of fusible numbers, which we call \emph{tame fusible numbers}.

$\ff'$ is defined from the bottom up by transfinite induction. At each stage of the induction, an initial segment $\mathcal H$ of the current $\ff'$ is marked as ``used''. The following invariant is kept throughout the construction process: If $x = \min(\ff'\setminus\mathcal H)$ is the smallest unused element, then $\ff'\cap[0,x+1)$ has already been defined, and $x+1$ is a limit point in this set.

\begin{figure}
\centerline{\includegraphics{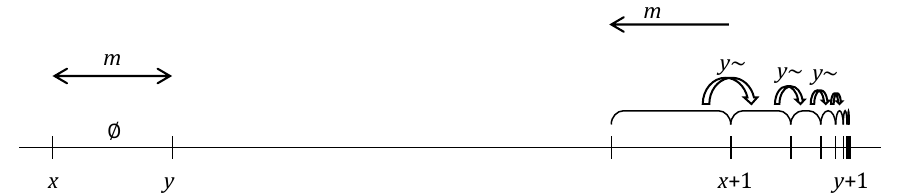}}
\caption{\label{fig_tame}Constructing the tame fusible numbers.}
\end{figure}

We start by leting $\ff'\cap [0,1)=\{1-2^{-n}:n\in\N\}$ and $\mathcal H=\emptyset$. Now let $x =\min (\ff'\setminus\mathcal H)$ be the smallest unused element, and suppose by induction that $\ff'\cap[0,x+1)$ has already been defined, and that $x+1$ is a limit point in this set. Let $y$ be the successor of $x$ in $\ff'$, and let $m=y-x$. We define $\ff'\cap [x+1,y+1)$ by taking the interval $\ff'\cap I_0$ for $I_0 = [x+1-m,x+1)$ (which has already been defined), and applying to it ``$y\fuse{}$'', obtaining $\ff'\cap I_1$ for $I_1 = [x+1,x+1+m/2)$, and then applying  ``$y\fuse{}$'' to this latter set to obtain $\ff'\cap I_2$ for $I_2 = [x+1+m/2,x+1+3m/4)$, and so on. In general, denoting $I_n = [y+1-m/2^{n-1},y+1-m/2^n)$ for $n\in\N$, each interval $\ff'\cap I_{n+1}$ is a scaled-down copy of $\ff'\cap I_n$ obtained by applying ``$y\fuse{}$''. For each $n\in\N$ let $\ell_n = y+1-m/2^{n-1}$ be the left endpoint of $I_n$. We next add $\ell_n$ for each $n\ge 1$ to $\ff'$, if these elements are not already there. (We will prove later on that this operation is unnecessary since we always have $\ell_0\in\ff'$.) Finally, we add $x$ to $\mathcal H$ and repeat.

At the end of the transfinite induction, $\ff'\cap[0,\infty)$ is defined, and $\mathcal H = \ff'$. See Figure~\ref{fig_tame}. The set $\ff'$ is closed by construction, thanks to the addition the endpoints $\ell_n$ in the construction.

Not all fusible numbers are tame. The smallest ``wild'' fusible number that we know of is $8449/4096$.

\begin{table}
\begin{multicols}{2}
\begin{algorithm}[H]
\caption{Finds $\min(\ff'\cap(r,\infty))$}
\begin{algorithmic}[5]
\Procedure{TameSucc}{$r$}
\If{$r<0$}
\State\Return{$0$}
\EndIf
\State $x\gets$\Call{TameSucc}{$r-1$}
\State $y\gets$\Call{TameSucc}{$2r-x-1$}
\State\Return $x\fuse y$
\EndProcedure
\end{algorithmic}
\end{algorithm}

\begin{algorithm}[H]
\caption{Computes $\textsc{TameSucc}(r)-r$}
\begin{algorithmic}[5]
\Procedure{M}{$r$}
\If{$r<0$}
\State\Return{$-r$}
\EndIf
\State\Return{$\textsc{M}(r-\textsc{M}(r-1))/2$}
\EndProcedure
\end{algorithmic}
\end{algorithm}
\end{multicols}
\caption{\label{table_tame_algs}Algorithms for tame fusible numbers.}
\end{table}

The recursive algorithm $\TS(r)$ shown in Table~\ref{table_tame_algs} returns the smallest tame fusible number larger than $r$, and $\MT$ is a concise algorithm that returns the difference $\TS(r)-r$.

\begin{lem}\label{lem_lower_bd}
Given $z\in\ff'$, let $\alpha = \ord(\ff'\cap[0,z))$ and let $\beta = \ord(\ff'\cap[0,z+1))$. Then $\beta = \omega^{1+\alpha}$. (Hence, for $z\ge 1$ we have $\beta = \omega^\alpha$.)
\end{lem}

\begin{proof}
By transfinite induction on $\alpha$. If $\alpha=0$ then $z=0$ and $\beta = \omega$. Next, suppose that $\alpha=\gamma+1$ is a successor ordinal, and let $y$ be the predecessor of $z$ in $\ff'$. Assume by induction that $\ord(\ff'\cap[0,y+1))= \omega^{1+\gamma}$. Let $m = z-y$. Consider the intervals $I_n = [z+1-2^{-(n-1)}m, z+1-2^{-n}m)$ for $n\in\N$. Let $\mathcal G = \ff\cap I_0$. We have $\ord(\mathcal G) = -\delta+\omega^\gamma$ for some $\delta<\omega^\gamma$, so $\ord(\mathcal G) = \omega^\gamma$. Since $[y+1,z+1)$ contains $\omega$-many copies of $\mathcal G$, we have
\begin{equation*}
\ord(\ff'\cap[0,z+1)) = \omega^{1+\gamma} + \omega^{1+\gamma}\omega = \omega^{1+\gamma+1} = \omega^{1+\alpha},
\end{equation*}
as desired.

Finally, suppose $\alpha$ is a limit ordinal. For every $\beta<\alpha$ there exists a corresponding fusible number $y<z$, which by induction satisfies $\ord(\ff'\cap[0,y+1)) = \omega^{1+\beta}$. Furthermore, since $\ff'$ is closed, as $\alpha\to\beta$ we have $y\to z$. Hence, $\ord(\ff'\cap[0,z+1))=\lim_{\beta<\alpha}\omega^{1+\beta} = \omega^{1+\alpha}$.
\end{proof}

Hence, for every positive integer $n$ we have $\ord(n)\ge \tau_n$, and $\ord(\ff')= \eps_0$, proving the lower bound of Theorem~\ref{thm_eps0}.

\subsection{Recurrence relations for the gaps}\label{sec_tame_rec}

Given $z\in\ff'$, define $\ord'(z) = 1 + \ord(\ff'\cap[0,z))$. Hence, by Lemma~\ref{lem_lower_bd} we have $\ord'(z+1) = \omega^{\ord'(z)}$ for all $z$, and for $z\ge 1$ the ``$1+{}$'' has no effect. Given $0<\alpha<\eps_0$, define $\fus'(\alpha)$ as the unique $z\in\ff'$ with $\ord'(z) = \alpha$, define $m(\alpha) = \fus'(\alpha+1) - \fus'(\alpha)$, and define $d(\alpha) = -\log_2 m(\alpha)$.

Hence, $d$ is a function from ordinals to natural numbers. Our objective is to show that $d(\alpha)$ experiences $F_{\eps_0}$-like growth. For this, we first need to derive recurrence relations for $d$. Unfortunately, the recurrence relations are quite complex, as they involve an auxiliary function $\exc(\alpha)$, which is itself defined recursively in terms of $\exc$ and $d$.

\begin{lem}\label{lem_recexc}
Let $\gamma<\eps_0$ be a limit ordinal, and let $y=\fus'(\gamma) - m(\gamma)$. Then $y\in\ff'$, and furthermore, $\ord'(y) = [\gamma]_{\exc(\gamma)}$ for some integer $\exc(\gamma)$. Specifically, $\exc(\gamma)$ is given recursively as follows:
\begin{align}
\exc(\omega^\alpha(k+1)+\beta) &= \exc(\omega^\alpha+\beta),&&\text{$k\ge 1$, $\beta<\omega^\alpha$ limit};\label{eq_exc1}\\
\exc(\omega^\alpha+\beta) &= \exc(\omega^{[\alpha]_{\exc(\alpha)}}+\beta),\footnotemark&&\text{$\alpha, \beta$ limits, $\beta<\omega^\alpha$, $\beta\neq\omega^{[\alpha]_{\exc(\alpha)}+1}$};\label{eq_exc2}\\
\exc(\omega^\alpha + \omega^{[\alpha]_{\exc(\alpha)}+1}) &= \exc(\omega^{[\alpha]_{\exc(\alpha)}+1})-1,&&\text{$\alpha$ limit};\label{eq_exc3}\\
\exc(\omega^{\alpha+1}+\beta) &= \exc(\omega^\alpha 2 + \beta),&&\text{$\beta<\omega^{\alpha+1}$ limit};\label{eq_exc4}\\
\exc(\omega^{\alpha+1}(k+1))&= \exc(\omega^{\alpha+1})-2,&&\text{$k\ge 1$};\label{eq_exc5}\\
\exc(\omega^{\alpha+1})&=d(\omega^{\alpha+1}) - d(\alpha) + 1;\label{eq_exc6}\\
\exc(\omega^\alpha k) &= d(\omega^\alpha) - d(\alpha) + \exc(\alpha), &&\text{$k\ge 1$, $\alpha$ limit.}\label{eq_exc7}
\end{align}
\footnotetext{Note that if $\beta$ is large enough then $\omega^{[\alpha]_{\exc(\alpha)}}+\beta=\beta$.}
Moreover for every $n\ge 0$ we have $\fus'([\gamma]_{\exc(\gamma)+n}) = \fus'(\gamma)- m(\gamma)/2^n$.
\end{lem}

\begin{figure}
\centerline{\includegraphics{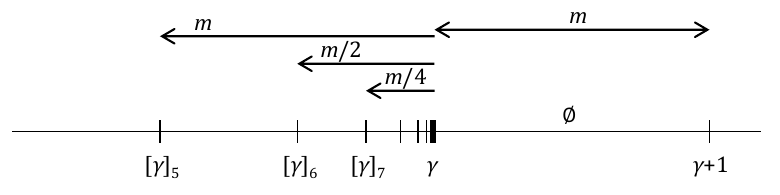}}
\caption{\label{fig_excess}Illustration for Lemma~\ref{lem_recexc}. We have $\exc(\gamma)=5$ in this example.}
\end{figure}

See Figure~\ref{fig_excess}. Lemma~\ref{lem_recexc} implies that, in the construction of $\ff'$, we always have $\ell_0\in\ff'$, so the addition of the numbers $\ell_n$, $n\ge 1$ to $\ff'$ was unnecessary.

\begin{proof}
By ordinal induction on $\gamma$. Let $\gamma = \omega^{\alpha}k+\beta$ be a limit ordinal, with $\beta<\omega^\alpha$. Let $y=\fus'(\gamma)$ and $z=\fus'(\gamma+1)$. Let $\ell_n = \fus'(\alpha+1)+1-2^{1-n}m(\alpha)$ for $n\in\N$, so $I_n = [\ell_n,\ell_{n+1})$ were the intervals used in the construction of $\ff'$. Then $y\in I_k$. Hence, for $n\ge 1$ we have $\ord'(\ell_n) = \omega^{\alpha} n$.

What is $\ord'(\ell_0)$? If $\alpha = \alpha'+1$ is a successor ordinal, then $\ord'(\ell_0) = \omega^{\alpha'}2$. And if $\alpha$ is a limit ordinal, then letting $z_0 = \ell_0-1$, $z_1 = \ell_1-1$, we have $\ord'(z_1) = \alpha$, so by the induction assumption, $\ord'(z_0) = [\alpha]_{\exc(\alpha)}$, and hence $\ord'(\ell_0) = \omega^{[\alpha]_{\exc(\alpha)}}$.

Each $\ff'\cap I_n$ is a scaled copy of $\ff'\cap I_{n-1}$. Hence, whenever $u\in\ff'\cap I_n$ for some $n\ge 1$, if $u'\in \ff'\cap I_{n-1}$ is the pre-image from which $u$ was copied (so $u=\fus'(\alpha+1)\fuse u'$), then the relation between $\ord'(u')$ and $\ord'(u)$ is
\begin{equation}\label{eq_useminus}
-\ord'(\ell_{n-1}) + \ord'(u') = - \ord'(\ell_n) + \ord'(u).
\end{equation}
In our case, let $y',z'\in\ff'\cap I_{k-1}$ be the images from which $y,z$ were copied.

Letting $\gamma'=\ord'(y')$, we have $\gamma' = \ord'(\ell_{k-1})+\beta$ by (\ref{eq_useminus}). Let $x'_n = \fus'([\gamma']_{\exc(\gamma')+n})$ for each $n\in\N$. By induction assumption we have $x'_n = y' - m(y')/2^n$.

Suppose first that $\beta>0$. In this case, for every $p\ge 2$ we have $[\gamma']_p > \ord'(\ell_{k-1})$ (the case in which $k=1$ and $\alpha$ is a limit ordinal follows from the Bachmann property; all the remaining cases are simpler). Hence, $x'_n \in I_{k-1}$ for all $n$.

For each $n$ let $x_n\in I_k$ be the image of $x'_n$. By (\ref{eq_useminus}) we have
\begin{equation*}
\ord'(x_n) = \ord'(\ell_n) + (-\ord'(\ell_{n-1})+\ord'(x'_n)).
\end{equation*}
Furthermore, the ratios between distances among $x'_n,y',z'$ are preserved among $x_n,y,z$. Hence, (\ref{eq_exc1}--\ref{eq_exc4}) follow, using Lemma~\ref{lem_hiccup} for (\ref{eq_exc3}).

Now suppose $\beta=0$, so $\gamma = \omega^\alpha k = \ord'(\ell_k)$. In this case the numbers $x_n$ belong to $I_{k-1}$. If $k\ge 2$ then $I_{k-1}\cup I_k$ is a scaled copy of $I_{k-2}\cup I_{k-1}$, and (\ref{eq_exc5}) follows.

Now suppose $k=1$, so $\gamma = \omega^\alpha$. If $\alpha = \alpha'+1$ is a successor ordinal, then for every $p\ge 1$ we have
\begin{equation*}
\fus'([\gamma]_p) = \fus'(\omega^{\alpha'}p) = y- 2^{1-p} m(\alpha') = y - 2^{1-p-d(\alpha')}.
\end{equation*}
Hence, $\fus'(\gamma) - m(\gamma)/2^n = y - 2^{-d(\gamma)-n} = \fus'([\gamma]_p)$ for $p = (d(\gamma) - d(\alpha')+1)+n$, yielding (\ref{eq_exc6}).

Finally, if $\gamma=\omega^\alpha$ for limit $\alpha$, then $\fus'(\alpha)=y-1$. We know by induction that
\begin{equation*}
\fus'([\alpha]_{\exc(\alpha) + p}) = y-1-m(\alpha)/2^p = y-1-2^{-d(\alpha)-p}\qquad\text{for every $p\in\N$.}
\end{equation*}
Hence, for every $q\ge \exc(\alpha)$ we have $\fus'([\omega^\alpha]_q) = y-2^{-d(\alpha)-q + \exc(\alpha)}$. Hence, $y-m(\gamma)/2^n = y - 2^{-d(\gamma)-n}$ is obtained by taking $q = (d(\gamma) - d(\alpha)+\exc(\alpha))+n$, yielding (\ref{eq_exc7}).
\end{proof}

\begin{lem}\label{lem_recd}
Let $0<\gamma<\eps_0$. Then $d(\gamma)$ is given recursively by:
\begin{align}
d(1) &= 1;\\
d(\alpha+1) &= 1+ d(\alpha);\label{eq_d1}\\
d(\omega^\alpha\cdot (k+1) + \beta) &= k + d(\omega^\alpha+\beta),&&\beta<\omega^\alpha;\label{eq_d2}\\
d(\omega^{\alpha+1}+\beta) &= 1 + d(\omega^\alpha 2+\beta),&&\beta<\omega^{\alpha+1};\label{eq_d3}\\
d(\omega^\alpha+\beta) &= 1 + d(\omega^{[\alpha]_{\exc(\alpha)}}+\beta),&&\text{$\alpha$ limit, $\beta<\omega^\alpha$}.\label{eq_d4}
\end{align}
\end{lem}

\begin{proof}
Using similar considerations as in the proof of Lemma~\ref{lem_recexc}.
\end{proof}

Let us illustrate Lemmas~\ref{lem_recexc} and~\ref{lem_recd} by sketching the computation of $d(\omega^{\omega^\omega})$. We have $d(\omega^\omega) = 10$ and $\exc(\omega^\omega) = 11$. More generally,
\begin{equation*}
d\left(\omega^{\omega^n}\right) = n+9 + 4c_1+5c_2+\cdots + (n+2)c_{n-1},\qquad\text{for $n\ge 1$},
\end{equation*}
where $c_0 = 2$, and $c_n = 1 +(n+1)c_{n-1}$ for $n\ge 1$. Hence, $d\left(\omega^{\omega^n}\right)$ grows with $n$ roughly like $n!$. The precise value of $d(\omega^{\omega^\omega})$ is $d(\omega^{\omega^\omega})=1+d(\omega^{\omega^{11}}) = 1541023937$.

\subsection{Growth rate}\label{sec_growth_rate}

We now prove that there exists a constant $c$ such that $d(\tau_n) \ge F_{\eps_0}(n-c)$ for all large enough $n$. We achieve this by bounding $d$ from below in terms of the Hardy hierarchy, since it is easier to work with.

\begin{lem}
For every $\alpha$ we have $d(\omega^\alpha) > d(\alpha)$.
\end{lem}

\begin{proof}
This is immediate from the construction of $\ff'$.
\end{proof}

\begin{cor}
For every limit $\alpha$ we have $\exc(\omega^\alpha) > \exc(\alpha)$.
\end{cor}

\begin{proof}
By (\ref{eq_exc7}).
\end{proof}

\begin{obs}\label{obs_ddescent}
Starting from an expression of the form $d(\omega^\gamma)$, repeated application of (\ref{eq_d2}--\ref{eq_d4}) yields expressions of the form $n_k+d(\omega^{\gamma_k})$, where $\gamma_1,\gamma_2,\ldots$ is a canonical descent starting from $\gamma$, each $n_{k+1} \in \{n_k+1,n_k+2\}$, and the descent parameter at step $k$ is $\exc(\gamma_{k-1})$.
\end{obs}

\begin{lem}
For every $\alpha$ we have $d\bigl(\omega^{\omega^\alpha}\bigr) \ge 2d(\omega^\alpha)$.
\end{lem}

\begin{proof}
By Observation~\ref{obs_ddescent} and ordinal induction on $\alpha$. Suppose first that $\alpha$ is a limit ordinal. Then by (\ref{eq_d4}) we have $d(\omega^\alpha) = 1 + d(\omega^{[\alpha]_{\exc(\alpha)}})$, whereas $d\bigl(\omega^{\omega^\alpha}\bigr) = 1 + d\bigl(\omega^{\omega^{[\alpha]_n}}\bigr)$ for $n = \exc(\omega^\alpha)>\exc(\alpha)$. By the Bachmann property (Lemma~\ref{lem_nodetour}), the canonical descent from $\omega^{[\alpha]_n}$ must go through $\omega^{[\alpha]_{\exc(\alpha)}}$. Hence, using induction on $\alpha$,
\begin{equation*}
d\bigl(\omega^{\omega^\alpha}\bigr) \ge 2 + d\bigl(\omega^{\omega^{[\alpha]_{\exc(\alpha)}}}\bigr) \ge 2 + 2d(\omega^{[\alpha]_{\exc(\alpha)}}) = 2d(\omega^\alpha).
\end{equation*}
Similarly, for a successor ordinal, denoting $n = \exc(\omega^{\alpha+1})$,
\begin{equation*}
d\bigl(\omega^{\omega^{\alpha+1}}\bigr) = 1 + d\bigl(\omega^{\omega^\alpha n}\bigr) \ge 4 + d\bigl(\omega^{\omega^\alpha}\bigr) \ge 4 + 2d(\omega^\alpha) = 2d(\omega^{\alpha+1}),
\end{equation*}
since the canonical descent from $\omega^\alpha n$ must go through $\omega^\alpha$, and it must take at least three steps.
\end{proof}

\begin{cor}
For every $\alpha$ we have $\exc\bigl(\omega^{\omega^\alpha}\bigr) \ge d(\omega^\alpha)$.
\end{cor}

\begin{proof}
Substituting into (\ref{eq_exc7}).
\end{proof}

\begin{lem}
Define
\begin{equation*}
f_{\beta,\alpha}(n) = d\Bigl(\omega^{\omega^{\omega^{\beta+\omega^\alpha n}}}\Bigr).
\end{equation*}
Then $f_{\beta,\alpha}(n) \ge H_\alpha(n)$ for all $\beta$.
\end{lem}

\begin{proof}
By ordinal induction on $\alpha$. For $\alpha=0$ we have
\begin{equation*}
f_{\beta,0}(n) = d\Bigl(\omega^{\omega^{\omega^{\beta + n}}}\Bigr) \ge n = H_0(n),
\end{equation*}
by Observation~\ref{obs_ddescent}.

Next, suppose the claim is true for $\alpha$. Then
\begin{equation*}
f_{\beta,\alpha+1}(n) = d\Bigl(\omega^{\omega^{\omega^{\beta + \omega^{\alpha+1} n}}}\Bigr) = 1 + d\Bigl(\omega^{\omega^{\omega^{\beta + \omega^{\alpha+1}(n-1) + \omega^\alpha m}}}\Bigr),
\end{equation*}
where
\begin{equation*}
m = \exc\bigl(\omega^{\omega^{\beta + \omega^{\alpha+1} n}}\bigr) \ge d\bigl(\omega^{\beta + \omega^{\alpha+1} n}\bigr) \ge n+1.
\end{equation*}
Hence, letting $\beta'=\beta + \omega^{\alpha+1}(n-1) + \omega^\alpha(m-n-1)$,
\begin{equation*}
f_{\beta,\alpha+1}(n) \ge d\Bigl(\omega^{\omega^{\omega^{\beta'+\omega^\alpha(n+1)}}}\Bigr) = f_{\beta',\alpha}(n+1) \ge H_\alpha(n+1) = H_{\alpha+1}(n).
\end{equation*}

Finally, let $\alpha$ be a limit ordinal, and suppose the claim is true for all $\alpha'<\alpha$. Then
\begin{equation*}
f_{\beta,\alpha}(n) = d\Bigl(\omega^{\omega^{\omega^{\beta + \omega^\alpha n}}}\Bigr) = 1 + d\Bigl(\omega^{\omega^{\omega^{\beta + \omega^\alpha(n-1) + \omega^{[\alpha]_m}}}}\Bigr),
\end{equation*}
where $m\ge n+1$ as before. By the Bachmann property, any canonical descent from $[\alpha]_m$ must go through $[\alpha]_n+1$. Hence,
\begin{equation*}
f_{\beta,\alpha}(n) \ge d\Bigl(\omega^{\omega^{\omega^{\beta +\omega^\alpha(n-1) + \omega^{[\alpha]_n+1}}}}\Bigr) = 1 + d\Bigl(\omega^{\omega^{\omega^{\beta +\omega^{\alpha}(n-1) + \omega^{[\alpha]_n}p}}}\Bigr),
\end{equation*}
where $p\ge n$ as before. Hence, letting $\beta' = \beta+\omega^\alpha(n-1) + \omega^{[\alpha]_n}(p-n)$, we have
\begin{equation*}
f_{\beta,\alpha}(n) \ge d\Bigl(\omega^{\omega^{\omega^{\beta' + \omega^{[\alpha]_n}n}}}\Bigr) = f_{\beta',[\alpha]_n}(n) \ge H_{[\alpha]_n}(n) = H_\alpha(n).\qedhere
\end{equation*}
\end{proof}

\begin{cor}\label{cor_d_fast}
For every $n\ge 8$ we have $d(\tau_n)\ge F_{\eps_0}(n-7)$.
\end{cor}

\begin{proof}
Any canonical descent from $\tau_{n-1}$ with descent parameter at least $2$ must go through $\gamma = \omega^{\omega^\delta}$ for $\delta = \tau_{n-4}2$. Hence, by Observation~\ref{obs_ddescent} and Lemma~\ref{lem_columnize_eps0} we have
\begin{equation*}
d(\tau_n) =d(\omega^{\tau_{n-1}})\ge d(\omega^\gamma) = f_{0,\tau_{n-5}}(2) \ge H_{\tau_{n-5}}(2) \ge F_{\eps_0}(n-7).\qedhere
\end{equation*}
\end{proof}

Finally, we prove that for every $\alpha\ge \tau_n$ we have $d(\alpha) \ge d(\tau_n)$.

\begin{lem}\label{lem_more_terms}
If $0<\beta < \omega^{\alpha+1}$ then $d(\omega^\alpha) < d(\omega^\alpha + \beta)$.
\end{lem}

\begin{proof}
By ordinal induction on $\gamma=\omega^\alpha+\beta$. Suppose first that $\alpha$ is a limit ordinal. Denote $k=\exc(\alpha)$, so $d(\omega^\alpha) = 1 + d(\omega^{[\alpha]_k})$ by (\ref{eq_d4}). Let $\omega^\delta$ be the first term of $\beta$. If $\delta=\alpha$ then by (\ref{eq_d2}) we have $d(\gamma) > d(\omega^\alpha + \beta')$ for some $\beta'<\beta$, so the claim follows by induction.

Now suppose $\delta<\alpha$, so $d(\gamma) = 1 + d(\omega^{[\alpha]_k} + \beta)$ by (\ref{eq_d4}). If $\beta < \omega^{[\alpha]_k+1}$ then the claim follows by induction. Otherwise we have $\omega^{[\alpha]_k}+\beta=\beta$, and the first term of $\beta$ is $\omega^\delta$ for some  $[\alpha]_k < \delta \le \alpha$. By induction we have $d(\beta) \ge d(\omega^\delta)$, which by the Bachmann property and Observation~\ref{obs_ddescent} is larger than $d(\omega^{[\alpha]_k})$.

If $\alpha$ is a successor ordinal we proceed similarly.
\end{proof}

\begin{lem}\label{lem_tau_largest}
If $\alpha> \tau_n$ then $d(\alpha)> d(\tau_n)$.
\end{lem}

\begin{proof}
If the first term of $\alpha$ is $\tau_n$ then the claim follows by Lemma~\ref{lem_more_terms}. Otherwise, the first term of $\alpha$ is $\omega^\beta$ for some $\beta > \tau_{n-1}$. By Lemma~\ref{lem_more_terms} we have $d(\alpha) \ge d(\omega^\beta)$. Furthermore, any canonical descent from $\beta$ must contain $\tau_{n-1}$, so the claim follows by Observation~\ref{obs_ddescent}.
\end{proof}

Corollary~\ref{cor_d_fast} together with Lemma~\ref{lem_tau_largest} implies Theorem~\ref{thm_fast}. This in turn implies a similar lower bound for the running time of the algorithms $\textsc{TameSucc}$ and $\succe$, since in order to output a fraction with denominator $2^d$, the recursion depth must be at least $d$.

\section{Upper bound}\label{sec_upper}

Here we show that $\ord(\ff)\le \eps_0$.

\begin{lem}
Let $\mathcal G_1, \mathcal G_2\subset \R$ be two well-ordered sets, and let $\mathcal H = \{x\fuse y : x\in\mathcal G_1, y\in\mathcal G_2, |y-x|<1\}$. Then $\ord(\mathcal H)\le \ord(\mathcal G_1)\otimes\ord(\mathcal G_2)$.
\end{lem}

\begin{proof}
By the above-mentioned result of de Jongh and Parikh~\cite{dJP}.
\end{proof}

\begin{lem}\label{lem_upper_step}
Let $y,z\in\ff$ satisfy $z=s(y)$. Suppose $\ord(y+1)\le \omega^{\omega^\alpha}$. Then $\ord(z+1) \le \omega^{\omega^{\alpha+1}}$.
\end{lem}

\begin{proof}
Let $J_1 = [0,y+1)$ and $J_2 = [y+1,z+1)$. Let $m=z-y$. Define the numbers $\ell_n = z+1-2^{-(n-1)}m$ and the intervals $I_n = [\ell_n, \ell_{n+1})$ for $n\in\N$. Hence, $\ell_1=y+1$, so the interval $I_0$ lies at the end of $J_1$, while the intervals $I_1,I_2,I_3,\ldots$ disjointly cover $J_2$. Given $n\in\N$, let $\ff_n=\ff\cap I_n$ and $\mathcal G_n = \ff\cap [0,\ell_{n+1})$, so $\mathcal G_{n+1} = \mathcal G_n\cup \ff_{n+1}$. We will bound $\ord(\mathcal G_n)$ by induction on $n$.

There is no fusible number between $y$ and $z$, and already ``$z\fuse{}$'' sends $\ell_n$ into $\ell_{n+1}$. Therefore, every $x\in\ff_n$, $n\ge 1$ must be of the form $x=x_1\fuse x_2$ for $x_1,x_2<\ell_n$. Therefore,
\begin{equation*}
\ord(\ff_n)\le \ord(\mathcal G_{n-1})\otimes\ord(\mathcal G_{n-1}).
\end{equation*}
Furthermore, $\ord(\mathcal G_{n+1}) = \ord(\mathcal G_n)+\ord(\ff_{n+1})$. Hence, it follows by induction on $n$ that $\ord(\ff_n)$, $\ord(\mathcal G_n)$ are both bounded by $\omega^{\omega^\alpha 2^n}$ for every $n\ge 1$.

Since $\omega^{\omega^{\alpha+1}} > \omega^{\omega^\alpha m}$ for every $m\in\N$, the claim follows.
\end{proof}

\begin{cor}
For every $z\in\ff$ we have $\ord(z+1)\le \omega^{\omega^{\ord(z)}}$.
\end{cor}

\begin{proof}
By transfinite induction on $z$, as in the proof of Lemma~\ref{lem_lower_bd}.
\end{proof}

Hence, for every positive integer $n$ we have $\ord(n)\le \tau_{2n-1}$, and $\ord(\ff)\le \eps_0$, proving the upper bound of Theorem~\ref{thm_eps0}.

\subsection{A shorter proof of the upper bound}

As one of the referees pointed out, there is a shorter way to prove that $\ord(\ff)\le \eps_0$, by invoking Kruskal's tree theorem~\cite{kruskal} and the upper bounds on the maximal order types of the corresponding well-partial-orders~\cite{hasegawa,RW93,Sch79,Sch20}. Let $T_1$ and $T_2$ be the binary-tree representations of two fusible numbers. It can be easily checked that, if $T_1$ embeds into $T_2$ in the sense of Kruskal's tree theorem, then $v(T_1)\le v(T_2)$. Hence, $\ff$ is a linearization of a subset of the well-partially-ordered set of unlabeled binary trees. Therefore, $\ord(\ff)\le \eps_0$ (as can be seen e.g.~from Theorem 3.2 of~\cite{Sch20}, by taking $T(X,\emptyset,X)$ for a one-element set $X$, and noting that $f^+\binom{0\,0\,1}{0\,1\,2}=f\binom{0\,0\,1}{0\,1\,2}=f\binom{1}{2}=\eps_0$).

(For a similar appeal to Kruskal's tree theorem, see e.g.~Ehrenfeucht~\cite{ehrenfeucht}.)

\section{Results on Peano Arithmetic}\label{sec_PA}

\subsection{Background}

(For more details see e.g.~Rautenberg~\cite{rautenberg}.) \emph{First-order arithmetic} is a formal language that includes the quantifiers $\exists$, $\forall$; logical operators $\wedge$, $\vee$, $\neg$; the constant symbol $0$; variables $x,y,z,\ldots$ which represent natural numbers; the equality and inequality symbols $=$, $\neq$; the binary operators $+$, $\cdot$; and the unary operator $S$, which represents the operation of adding $1$. The operator $S$ together with the symbol $0$ allows us to express any natural number. For example, $3$ can be represented by $SSS0$. However, for readability we will use the notation ``$3$''.

Any arithmetical formula with no free variables is either true or false. For example, the formula $\forall x\, \exists y\, x+2=y$ is true, whereas the formula $\forall x\, \exists y\, y+2=x$ is false since the variables range over $\N$.

One can express the notion ``$x<y$'' by the formula $\exists z\, x+z+1=y$, or else by the formula $\forall z\, y+z\neq x$. The formula
\begin{equation*}
x\bmod y = z \defisymbol z<y\wedge\exists q\, x=q\cdot y+z
\end{equation*}
expresses the remainder operation as a ternary relation on the free variables $x$, $y$, $z$.

One can encode any pair of numbers as a single number by exploiting the fact that the \emph{pairing function} $\langle a,b\rangle = (a+b)(a+b+1)/2+a$ is a bijection between $\N^2$ and $\N$. Hence, the formula
\begin{equation*}
\langle a,b\rangle = c \defisymbol 2\cdot c=(a+b)\cdot(a+b+1)+2\cdot a
\end{equation*}
expresses the pairing function as a ternary relation. By repeatedly applying the pairing function we can encode any $k$-tuple as a single natural number. The number $k$ must be fixed in advance, however. Hence, it is not immediately clear how to encode sequences of \emph{arbitrary} length in a way that enables access to each element of the sequence. G\"odel found a way to accomplish this using the Chinese Remainder Theorem. The idea is to encode the sequence $a_0, \ldots, a_{n-1}$ as $a_i = x\bmod r_i$ where the numbers $r_i$ form an arithmetic progression. Specifically, one can take $r_i = (i+1)m+1$ for some appropriately chosen $m$. We can also provide the length of the sequence as an additional parameter. Hence, the number $z = \langle \langle n,x\rangle,m\rangle$ encodes the finite sequence $a_0, \ldots, a_{n-1}$ in which $a_i = x\bmod ((i+1)m+1)$ for each $0\le i<n$. Denote by $[z]_i = a$ the statement that ``the $i$-th element of the sequence encoded by $z$ equals $a$''. Then we can express ``$[z]_i = a$'' in first-order arithmetic as a ternary relation:
\begin{equation*}
[z]_i = a \defisymbol \exists q\, \exists n\, \exists x\, \exists m\, \langle n,x\rangle = q\wedge \langle q,m\rangle = z\wedge i<n\wedge x\bmod ((i+1)\cdot m+1)=a.
\end{equation*}

Using finite sequences, it is possible to represent exponentiation, factorials, Fibonacci numbers, etc. We can also represent any kind of discrete structures, such as trees, graphs, etc. We can also represent rational numbers and their arithmetical operations.

Once we have a representation of rational numbers, we can formulate the statement ``the rational number $x$ is fusible" as follows: ``There exists a finite set $S$ of rational numbers that includes $x$, and such that for every $w \in S$, either $w=0$ or there exist $y,z \in S$ such that $|z-y|<1$ and $w=y\fuse z$." And we can formulate the statement ``algorithm $\TS$ terminates on input $r_0$ outputting $f_0$'' as follows: ``There exists a finite set $S$ of quadruples $(r,x,y,f)$ that contains $(r_0,x,y,f_0)$ for some $x,y$, and such that for every $(r,x,y,f) \in S$, if $r<0$ then $f=0$, otherwise $f=x\fuse y$ and we have $(r-1,x',y',x),(2r-x-1,x'',y'',y)\in S$ for some $x',y',x'',y''$.''

More generally, in first-order arithmetic we can represent Turing machines and formulate statements such as ``Turing machine $T$ halts on input $x$".

However, in first-order arithmetic it is not possible to quantify over all real numbers, nor over all infinite sequences or sets, as follows trivially from countability considerations (though one can formulate statements about specific real numbers such as $\pi$ and $e$). Still, one can quantify over infinite sequences in which each element of the sequence is uniquely determined in some pre-defined way by the previous elements.

\emph{Peano Arithmetic} is a theory for first-order arithmetic given by a set of axioms. The \emph{PA-theorems} are those statements that can be derived from the axioms and from previously derived PA-theorems by the inference rules of first-order logic. Since the natural numbers satisfy the axioms, all PA-theorems are true (when interpreted in the natural numbers, which is the intended interpretation). Moreover, by use of the inference rules, any ``finitary'' mathematical proof (involving only finite sets and structures) can be carried out step by step in PA. However, as G\"odel showed, PA, like any formal system strong enough to represent basic arithmetic, is incomplete, in the sense that there exist true arithmetical statements that are not PA-theorems.

\paragraph{Unprovability in PA} There are finitary statements whose only obvious proofs involve infinitary arguments. Consider for example the following statement:
\begin{equation}\label{eq_descent_statement}
\begin{split}
&\text{For every ordinal $\alpha<\eps_0$, the canonical descent $\{\alpha_n\}$ given}\\&\text{by $\alpha_1=\alpha$ and $\alpha_{n+1} = [\alpha_n]_n$ eventually reaches $0$.}
\end{split}
\end{equation}
This statement can be formulated in first-order arithmetic. However, the obvious way to prove it is by arguing that the ordinals up to $\eps_0$ are well-ordered, so, more generally, there exists no infinite descending sequence starting from $\alpha$. Unfortunately, this latter statement involves quantification over infinite sequences. Hence, this proof approach to (\ref{eq_descent_statement}) cannot be carried out in PA.

An important tool for showing true arithmetical formulas to be unprovable in PA is the following result:

\begin{thmC}[\cite{BW}]\label{thm_BW}
Let $T$ be a Turing machine that computes a function $g:\N\to\N$, terminating on every input. Suppose that PA can prove the statement ``$T$ terminates on every input.'' Then $g$ cannot grow too fast: There exist $\alpha<\eps_0$ and $n_0\in\N$ such that $g(n) < F_\alpha(n)$ for every $n\ge n_0$.
\end{thmC}

Theorem~\ref{thm_BW} is due independently to Schwichtenberg~\cite{schwichtenberg} and Wainer~\cite{wainer1970,wainer1972}, building on earlier work by Kreisel~\cite{kreisel}. The above form follows Buchholz and Wainer~\cite{BW}, who give a nice, self-contained proof (which is actually a simple base-case of a much more general result of Buchholz~\cite{buchholz1987}).

Theorem~\ref{thm_BW} implies that statement (\ref{eq_descent_statement}) has \emph{no} proof in PA, because the length of the canonical descent starting from $\alpha=\tau_n$ grows like $F_{\eps_0}(n-c)$ for some $c$ (as quickly follows by comparing to the definition of $H_\alpha(n)$). Hence, let $T$ be a Turing machine that, given $n$, outputs the length of the canonical descent starting from $\tau_n$. If PA could prove (\ref{eq_descent_statement}), it could also conclude that $T$ halts on all inputs, contradicting Theorem~\ref{thm_BW}.

\subsection{Proof of Theorem~\ref{thm_PA}}

We start by analyzing algorithms $\TS$ and $\MT$ within PA. For convenience, in this section we will denote the algorithm $\TS$ by $\TSs$.

\begin{lem}
PA cannot prove the statement ``for every $n\in\N$, algorithm $\MT(n)$ terminates,'' nor the statement ``for every $n\in\N$, algorithm $\TSs(n)$ terminates.''
\end{lem}

\begin{proof}
Let $A$ be a Turing machine that, given $n$, outputs $1/\MT(n)$. If PA could prove that $\MT$ terminates, it could conclude that $A$ terminates as well. But this contradicts Theorem~\ref{thm_BW}, in light of Corollary~\ref{cor_d_fast}. The argument for $\TSs$ is similar.
\end{proof}

\begin{lem}\label{lem_TS_PA}
PA can prove that for every $r\in\Q$, if $\TSs(r)$ terminates then its output is a fusible number.
\end{lem}

\begin{proof}
We reason within PA. The general approach is to proceed by induction on the depth of the recursion. Suppose $\TSs(r)$ terminates. Recall that for $r<0$ the algorithm returns $0$, whereas for $r\ge 0$ the algorithm recursively sets $x=\TSs(r-1)$, $y=\TSs(2r-x-1)$, and then outputs $x\fuse y$. Denote $u=2r-x-1$ for convenience.

All we need to prove is that for $r\ge 0$ we have $|y-x|<1$, since then the claim follows immediately by induction. For $0\le r<1/2$ we have $x=y=0$. We will show that for $r\ge 1/2$ we in fact have $x<y<x+1$.

\begin{clm}\label{claim_r12}
If $\TSs(r)$ terminates then it returns $\TSs(r)>r$. If $r\ge -1/2$ then we also have $\TSs(r)\le r+1/2$.
\end{clm}

\begin{proof}
If $r<1/2$ the claim follows easily, so suppose $r\ge 1/2$. By induction we have $x=\TSs(r-1)\le r-1/2$, so $u\ge r-1/2\ge 0$, and by induction we have $u<y\le u+1/2$. Substituting this into $x\fuse y$ yields the desired result.
\end{proof}

Claim~\ref{claim_r12} implies that for $r\ge 1/2$ we have $x = \TSs(r-1) \le r - 1/2$, so $y=\TSs(2r-x-1) > 2r-x-1 \ge r-1/2\ge x$.

\begin{clm}\label{claim_joint}
Suppose $\TSs(r)$ terminates. Then:
\begin{enumerate}[label={\rm(\alph*)}]
\item If $r\ge 0$ then $y<x+1$.
\item For every $r<r'<\TSs(r)$, $\TSs(r')$ also terminates, outputting $\TSs(r') = \TSs(r)$. (Therefore, for every $r'>r$, if $\TSs(r')$ terminates then $\TSs(r')\ge \TSs(r)$).
\end{enumerate}
\end{clm}

\begin{proof}
We prove both properties jointly by induction on the recursion depth. Let us start with item (a). If $0\le r<1/2$ then $x=0$, $y=0$ and we are done. Hence, suppose $r\ge 1/2$. Claim~\ref{claim_r12} implies $r-1<x\le r-1/2$, so $0\le u<r$. We know $\TSs(u-1)$ terminates and $u-1<r-1$, so by induction, item (b) implies $\TSs(u-1)\le \TSs(r-1) = x$. The second recursive call in the computation of $T(u)$ is $v=\TSs(2u-\TSs(u-1)-1)$. By induction, item (a) implies $v<\TSs(u-1)+1$, so $y=\TSs(u) = \TSs(u-1)\fuse v < \TSs(u-1)\fuse (\TSs(u-1)+1) = \TSs(u-1)+1\le x+1$, $\TSs(r)=x\sim y<x+1=\TSs(r-1)+1$, and $\TSs(r)-1<\TSs(r-1)$.

For item (b), if $r<0$ the claim is immediate, so let $r\ge 0$ and let $r<r'<\TSs(r)$. Then $r-1<r'-1<\TSs(r)-1<\TSs(r-1)$, so by induction, item (b) implies that $\TSs(r'-1)$ terminates outputting $x$. Next, let $u'=2r'-x-1$. Then:
\begin{align*}
u'-u&=2(r'-r),\\
\TSs(u)-u &= 2(\TSs(r)-r).
\end{align*}
Therefore, $\TSs(u)-u' = 2(\TSs(r)-r')$, so $u'<\TSs(u)$. Hence, $u<u'<\TSs(u)$, so by induction, item (b) implies that $\TSs(u')$ terminates outputting $y$. Hence, $\TSs(r')$ terminates giving the same output as $\TSs(r)$.
\end{proof}

This finishes the proof of Lemma~\ref{lem_TS_PA}.
\end{proof}

\begin{lem}
PA can prove that for every $r\in\Q$, if $\TSs(r)$ does not terminate, then for every $n\in\N$ there exists a fusible number $z_n$ satisfying $r<z_n<r+2^{-n}$.
\end{lem}

\begin{proof}
We reason within PA. Suppose $\TSs(r)$ does not terminate. Then there exists an infinite recursive-call sequence with inputs $r_0, r_1, r_2, \ldots$ starting with $r_0=r$. Specifically, for each $n\ge 1$, \emph{either} $r_n=r_{n-1}-1$ and $\TSs(r_n)$ does not terminate, \emph{or} $\TSs(r_{n-1}-1)$ terminates outputting $x_n$, but $r_n = 2r_{n-1}-x_n-1$ and $\TSs(r_n)$ does not terminate. Either way, $r_n<r_{n-1}$.

The number of indices $n$ for which $r_n = r_{n-1}-1$ must be finite. Call this number $d$. We proceed by induction on $d$.

Suppose first that $d=0$, so $r_n = 2r_{n-1}-x_n-1$ for all $n$. Given $n\ge 1$, we know by Lemma~\ref{lem_TS_PA} that $x_n$ is fusible. Therefore, by the window lemma (Lemma~\ref{lem_window}) there exists a fusible number $z_n\in (r_n, r_{n-1}]$. Let $z_{n-1} = x_n\fuse z_n$. Then $z_{n-1}$ is fusible, and $r_{n-1}<z_{n-1} \le r_{n-2}$ (the lower bound follows from $z_n>r_n$, while the upper bound follows from $z_n\le r_{n-1}$ and $x_n\le x_{n-1}$, the latter of which follows from Claim~\ref{claim_joint} (b)). Furthermore, we have $z_n-r_n = 2(z_{n-1}-r_{n-1})$. In general, letting $z_i = x_{i+1}\fuse z_{i+1}$ for each $i=n-1,\ldots,1,0$, each $z_i$ is fusible and satisfies $r_i < z_i \le r_{i-1}$ and $z_i-r_i = 2(z_{i-1} - r_{i-1})$. But $z_n-r_n < r_{n-1} - r_n = x_n-(r_{n-1}-1) < 1/2$ by Claim~\ref{claim_r12}. Therefore, $z_0-r_0<2^{-n-1}$, as desired.

Now suppose the claim is true whenever the number of indices with $r_n=r_{n-1}-1$ is $d-1$, and suppose that in our sequence this number is $d$. Let $n$ be smallest such index in our sequence. By induction, for each $m$ there exists a fusible number $z_m$ satisfying $r_n < z_m < r_n+2^{-m}$. Then $z_m+1$ is also fusible, and if $m$ is large enough then it will fall in the range $(r_{n-1}, r_{n-2}]$. From here we proceed as before.
\end{proof}

Hence, PA cannot prove the statement ``for every $n\in\N$ there exists a smallest fusible number larger than $n$,'' because then PA would conclude by contradiction that $\TSs(n)$ terminates for all $n\in\N$.

Therefore, since PA can prove Lemma~\ref{lem_hmax_succ}, it follows that PA cannot prove the statement ``every fusible number has a maximum-height representation.''

\section{A conjecture}\label{sec_conj}

Given $x\in\ff$, let $y=s(x)$ be the successor of $x$, let $m = y-x$, let $\ell_n = y+1 - 2^{1-n}m$ for $n\in\N$, and let $I_{x,n} = [\ell_n,\ell_{n+1})$ for $n\in\N$, so the interval $I_{x,0}$ ends at $x+1$, while the intervals $I_{x,n}$, $n\ge 1$ form a partition of $[x+1,y+1)$. Recall that by Corollary~\ref{cor_next_omega} we have $s^{(n)}(x) = x+(2-2^{1-n})m$ for every $n\in\N$. We conjecture the following:

\begin{conj}\label{conj}
Every fusible number in $I_{x,n}$, $n\ge 1$ can be written as $s^{(n)}(x)\fuse z$ for some $z\in \ff\cap [x+1 - 2^{1-n}m,x+1)$.
\end{conj}

\begin{figure}
\centerline{\includegraphics{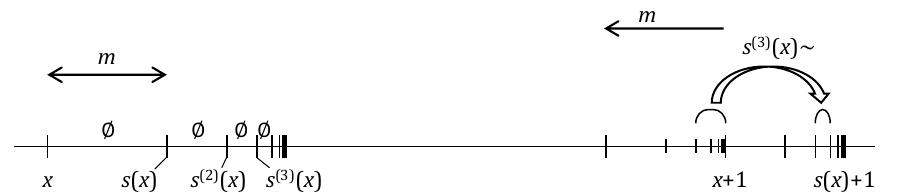}}
\caption{\label{fig_conj}We conjecture that all fusible numbers can be obtained this way.}
\end{figure}

\begin{table}
\begin{algorithm}[H]
\caption{Finds $\min(\ff''\cap(r,\infty))$}
\begin{algorithmic}[5]
\Procedure{MeekSucc}{$r$}
\If{$r<0$}
\State\Return{$0$}
\EndIf
\State $x\gets$\Call{MeekSucc}{$r-1$}
\State $y\gets$\Call{MeekSucc}{$2r-x-1-1/\textsc{Denominator}(x)+2^{\lceil\log_2(x-r+1)\rceil}$}
\State\Return $x\fuse y$
\EndProcedure
\end{algorithmic}
\end{algorithm}
\caption{\label{table_conj_alg}Algorithm for meek fusible numbers.}
\end{table}

See Figure~\ref{fig_conj}. In other words, by repeatedly performing the operations $s^{(n)}(x)\fuse z$, we obtain a subset $\ff''\subseteq\ff$, in a way analogous to the construction of the tame fusible numbers $\ff'$. Let us call the fusible numbers in $\ff''$ \emph{meek}.
 Conjecture~\ref{conj} states that $\ff''=\ff$. We have done extensive computer experiments, and we have not found any counterexample to Conjecture~\ref{conj}. Table~\ref{table_conj_alg} shows a recursive algorithm for the next meek fusible number, analogous to algorithm $\TS$ of Table~\ref{table_tame_algs}. If Conjecture~\ref{conj} is true, then this algorithm is equivalent to algorithm $\succe$ of Table~\ref{table_algs}.

\subsection{Recurrence relations for meek fusible numbers}

$\ff''$ satisfies recurrence relations similar to those of $\ff'$. Given $z\in\ff''$, define $\ord''(z) = 1 + \ord(\ff''\cap[0,z))$. Then $\ord''(z+1) = \omega^{\ord''(z)}$ for all $z$, and for $z\ge 1$ the ``$1+{}$'' has no effect. Given $0<\alpha<\eps_0$, define $\fus''(\alpha)$ as the unique $z\in\ff''$ with $\ord''(z) = \alpha$, define $m''(\alpha) = \fus''(\alpha+1) - \fus''(\alpha)$, and define $d''(\alpha) = -\log_2 m''(\alpha)$.

\begin{lem}\label{lem_recexcpp}
Let $\gamma<\eps_0$ be a limit ordinal, and let $y=\fus''(\gamma) - m''(\gamma)$. Then $y\in\ff''$, and furthermore, $\ord''(y) = [\gamma]_{\exc''(\gamma)}$ for some integer $\exc''(\gamma)$. Specifically, $\exc''(\gamma)$ is given recursively as follows:
\begin{align*}
\exc''(\omega^\alpha k+\beta) &= \exc''(\omega^{[\alpha]_{\exc''(\alpha)+k-1}}+\beta),&&\text{$k\ge 1$, $\alpha, \beta$ limits,}\\
&&&\qquad\text{$\beta<\omega^\alpha$ limit,}\\
&&&\qquad\text{$\beta\neq\omega^{[\alpha]_{\exc''(\alpha)+k-1}+1}$};\\
\exc''(\omega^\alpha k+\omega^{[\alpha]_{\exc''(\alpha)+k-1}+1}) &= \exc''(\omega^{[\alpha]_{\exc''(\alpha)+k-1}+1})-1,&&\text{$\alpha$ limit};\\
\exc''(\omega^{\alpha+1}k+\beta) &= \exc''(\omega^\alpha (k+1) + \beta),&&\text{$\beta<\omega^{\alpha+1}$ limit};\\
\exc''(\omega^\alpha k) &= d''(\omega^\alpha k) - d''(\alpha) + \exc''(\alpha) - 1,&&\text{$\alpha$ limit, $k\ge 2$};\\
\exc''(\omega^\alpha) &= d''(\omega^\alpha) - d''(\alpha) + \exc''(\alpha),&&\text{$\alpha$ limit};\\
\exc''(\omega^\alpha k) &= d''(\omega^\alpha k) - d''(\alpha) - k + 1,&&\text{$\alpha$ successor, $k\ge 2$};\\
\exc''(\omega^\alpha) &= d''(\omega^\alpha) - d''(\alpha) + 2,&&\text{$\alpha$ successor}.
\end{align*}
Moreover for every $n\ge 0$ we have $\fus''([\gamma]_{\exc''(\gamma)+n}) = \fus''(\gamma)- m''(\gamma)/2^n$. Furthermore, $d''(\gamma)$ is given recursively by:
\begin{align*}
d''(k) &= k;\\
d''(\omega^\alpha k+\beta) &= 1 + d''(\omega^{[\alpha]_{\exc''(\alpha)+k-1}} + \beta),&&\text{$k\ge 1$, $\alpha$ limit, $\beta<\omega^\alpha$};\\
d''(\omega^{\alpha+1} k+\beta) &= 1 + d''(\omega^\alpha(k+1)+\beta),&&\text{$k\ge 1$, $\beta < \omega^{\alpha+1}$}.
\end{align*}
\end{lem}

\subsection{Rate of growth of \texorpdfstring{$d''(\omega^{\omega^n})$}{d''(omega\^{}omega\^{}n)}}

We saw in Section~\ref{sec_tame_rec} that, for the tame fusible numbers, $d(\omega^{\omega^n})$ grows roughly like $n!$. In contrast, $d''(\omega^{\omega^n})$ exhibits Ackermannian growth, as we proceed to show.

First of all, we have:
\begin{align*}
d''(\omega^n k) &= 2n+k,\qquad k\ge 1;\\
\exc''(\omega^n) &=n+3;\\
\exc''(\omega^n k) &= n+1,\qquad k\ge 2;
\end{align*}
and, more generally, if $\alpha+\omega^n$ is in CNF with $0<\alpha<\omega^\omega$, then $\exc''(\alpha+\omega^n) = n+1$.

By repeatedly applying the recurrences for $d''$, starting from $d''(\omega^{\alpha_0}k_0)$, $\alpha_0<\omega^\omega$, we obtain
\begin{equation*}
d''(\omega^{\alpha_0}k_0)= 1 + d''(\omega^{\alpha_1} k_1)= 2 + d''(\omega^{\alpha_2} k_2)=\cdots
\end{equation*}
where the sequence $\{\alpha_n\}$ is a canonical descent.

Furthermore, the precise identities of the higher-order terms of $\alpha_0$ are unimportant for the descent until the lower-order terms disappear. Meaning, let $\alpha = \beta + \gamma$ and $\alpha' = \beta' + \gamma$ be two ordinals in CNF smaller than $\omega^\omega$, with the same lower-order terms $\gamma$, and with $\beta,\beta'>0$. Suppose that, starting from $d''(\omega^\alpha k)$, it takes $m$ steps to get down to $m + d''(\omega^\beta \ell)$.
Then, starting from $d''(\omega^{\alpha'} k)$, after the same number $m$ of steps we will get down to $m + d''(\omega^{\beta'} \ell)$ for the same $\ell$.

Hence, let $y_n(k)$ and $\ell_n(k)$ be the integers such that, starting from $d''(\omega^{\alpha+\omega^n}k)$, where $0<\alpha<\omega^\omega$ and $\alpha+\omega^n$ is in CNF, after $y_n(k)$ steps we get down to $y_n(k)+d''(\omega^\alpha\ell_n(k))$.

And more generally, let $y_{n,m}(k)$ and $\ell_{n,m}(k)$ be the integers such that, under the same conditions, starting from $d''(\omega^{\alpha+\omega^n m}k)$ we get down to $y_{n,m}(k) + d''(\omega^\alpha\ell_{n,m}(k))$.

Then,
\begin{align*}
y_{n,1}(k) &= y_n(k);\\
\ell_{n,1}(k) &= \ell_n(k);\\
y_{n,m}(k) &= y_n(k) + y_{n,m-1}(\ell_n(k)),\qquad m\ge 2;\\
\ell_{n,m}(k) &=\ell_{n,m-1}(\ell_n(k));\\
y_n(k) &= 1 + y_{n-1,n+k}(1),\qquad n\ge 2;\\
\ell_n(k) &= \ell_{n-1,n+k}(1),\qquad n\ge 2;
\end{align*}
with base cases $y_1(k) = \ell_1(k) = k+2$.

It follows that $\ell_{n,m}(k) = \ell_n^{(m)}(k)$ and $\ell_n(k) = \ell_{n-1}^{(n+k)}(1)$. Hence, $\ell_n(k)$ grows roughly like $F_{n-1}(k)$ in the fast-growing hierarchy. Further, $y_n(k)$ also grows roughly like $F_{n-1}(k)$.

Next, let $u_n(k)$ and $j_n(k)$ be such that
\begin{equation*}
    d''(\omega^{\omega^n}k) = u_n(k) + d''(\omega^{\omega^{n-1}}j_n(k)).
\end{equation*}
Then,
\begin{align*}
    u_n(k) &= 1 + y_{n-1,n+k+1}(1),\qquad n\ge 2;\\
    j_n(k) &= \ell_{n-1,n+k+1}(1),\qquad n\ge 2.
\end{align*}
Finally, let $v_n(k) = d''(\omega^{\omega^n}k)$. Then,
\begin{equation*}
    v_n(k) = u_n(k) + v_{n-1}(j_n(k)),\qquad n\ge 2;
\end{equation*}
with base case $v_1(k) = 2k+8$. Hence, $u_n(k)$, $j_n(k)$, and $v_n(k)$ also grow roughly like $F_{n-1}(k)$, and hence, $d''(\omega^{\omega^n})$ grows roughly like $F_\omega(n)$.

The precise value of $d''(\omega^{\omega^\omega})$ is given by $d''(\omega^{\omega^\omega}) = 1 + d''(\omega^{\omega^{11}}) = 1 + v_{11}(1)$, which is too large to calculate explicitly.

\section{Discussion}\label{sec_discussion}

A main open problem is to derive an upper bound for the inverse-largest-gap $g(n)^{-1}$. One way to do this would be to prove Conjecture~\ref{conj} and then establish an upper bound for the growth rate of $d''$.

We are also missing upper bounds for the running time of algorithms $\ISF$ and \textsc{WeakSucc}.

Experimentally, it seems that whenever a fusible number $z$ has representations $T_1$ and $T_2$ with heights $h(T_1)<h(T_2)$, it also has representations of all intermediate heights. However, we have not been able to prove this.

$\ACAzero$ is a subsystem of second-order arithmetic studied within the framework of reverse mathematics~\cite{simpson_sub}. Since $\ACAzero$ is a \emph{conservative extension} of PA (meaning that the first-order statements provable in $\ACAzero$ are exactly the PA-theorems), it follows that the second-order statements ``$\ff$ is well-ordered'', ``algorithm $\TS$ terminates on all real inputs'', and ``algorithm $\MT$ terminates on all real inputs'' are unprovable in $\ACAzero$.

\paragraph{Generalizations of fusible numbers} In a follow-up paper, the second author, together with Alexander Bufetov and Fedor Pakhomov, study sets of reals generated by $n$-ary functions~\cite{BNPinprep}. They show that the natural generalization of $\fuse$ to an $n$-ary function yields a well-ordered set of order type $\varphi_{n-1}(0)$. They also show that no linear $n$-ary function can get beyond $\varphi_{n-1}(0)$. Finally, they show that there exist continuous $n$-ary functions for which the order types of the resulting sets approach the small Veblen ordinal (matching the upper bound yielded by Kruskal's theorem).

\paragraph{Acknowledgements} Thanks to Fedor Pakhomov and the anonymous referees for their useful comments.

\bibliographystyle{alphaurl}
\bibliography{fusible_PA}

\end{document}